\documentclass[structabstract]{aa}  
\usepackage[pdftex]{graphics}
\usepackage{graphicx}
\usepackage{natbib}
\newcommand{\HI}{H\,{\sc i}}
\newcommand{\HII}{H\,{\sc ii}}
\newcommand{\Ha}{H$\alpha$}
\newcommand{\skms}{\ensuremath{\,\mbox{km}\,\mbox{s}^{-1}}}
\newcommand{\kms}{\ensuremath{\mbox{km}\,\mbox{s}^{-1}}}

\newcommand{\vsys}{\ensuremath{v_{\rm sys}}}

\newcommand{\vrot}{\ensuremath{v_{\rm rot}}}

\newcommand{\Msun}{~M$_{\odot}$}
\usepackage{txfonts}
\begin{document}
   \title{A kinematic study of the irregular dwarf galaxy NGC\,2366 using \HI\
   and \Ha\ observations}


   \author{J. van Eymeren
          \inst{1,2,3}
          \and 
          M. Marcelin\inst{4}
	  \and
	  B. Koribalski\inst{3}
	  \and
	  R.-J. Dettmar\inst{2}
	  \and
	  D.~J. Bomans\inst{2}
	  \and
	  J.-L. Gach\inst{4}
	  \and
	  P. Balard\inst{4}
          }

   \offprints{J. van Eymeren}

   \institute{Jodrell Bank Centre for Astrophysics, School of Physics \&
              Astronomy, The University of Manchester, Alan Turing Building,
              Oxford Road, Manchester, M13 9PL, UK\\
              \email{Janine.VanEymeren@manchester.ac.uk}
              \and
	      Astronomisches Institut der Ruhr-Universit\"at Bochum,
              Universit\"atsstra{\ss}e 150, 44780 Bochum, Germany
	      \and
	      Australia Telescope National Facility, CSIRO,
              P.O. Box 76, Epping, NSW 1710, Australia
              \and 
             Laboratoire d'Astrophysique de Marseille, OAMP, Universit\'e
              Aix-Marseille \& CNRS, 38 rue Fr\'ed\'eric Joliot-Curie, 13013
              Marseille, France
             }

   \date{Accepted 5 November 2008}

 
  \abstract
   {The metal content of dwarf galaxies and the metal enrichment of the
  intergalactic medium both suggest that mass loss from galaxies is a
  significant factor for the chemical evolution history of galaxies, in
  particular of dwarf galaxies. However, no clear evidence of a blow-away in
  local dwarf galaxies has been found so far.}
   {Dwarf galaxies provide a perfect environment for studying feedback
   processes because their kinematics and their generally low gravitational
   potential support the long-term survival of shells, filaments, and
   holes. We therefore performed a detailed kinematic analysis of the neutral
   and ionised gas in the nearby star-forming irregular dwarf galaxy NGC\,2366
   in order to make predictions about the fate of the gas and to get a more
   complete picture of this galaxy.}
   {A deep \Ha\ image and Fabry-Perot interferometric data of NGC\,2366 were
  obtained. They were complemented by \HI\ synthesis data from the THINGS
   survey. We searched for line-splitting both in \Ha\ and \HI\ by performing
   a Gaussian decomposition. To get an idea whether the expansion velocities
   are high enough for a gas blow-away, we used the pseudo-isothermal halo
   model, which gives us realistic values for the escape
   velocities of NGC\,2366. The good data quality also allowed us to discuss
   some peculiarities of the morphology and the dynamics in NGC\,2366.}
   {A large red-shifted outflow north west of the giant extragalactic \HII\
  region with an expansion velocity of up to 50\skms\ is found in \Ha, but not
  in \HI. Additionally, a blue-shifted component north of the
   giant extragalactic \HII\ region was detected both in \Ha\ and \HI\ with an
   expansion velocity of up to 30\skms. A comparison with the escape
   velocities of NGC\,2366 reveals that the gas does not have enough kinetic
   energy to leave the gravitational potential.}
   {This result is in good agreement with hydrodynamic simulations and suggests
  that we need to examine even less massive galaxies ($M\rm _{gas}=10^6\,M_{\odot}$).}


   \maketitle
%

\section{Introduction}
\label{Introduction}
\begin{figure*}
\centering
\includegraphics[width=\textwidth,viewport= 10 14 406 241]{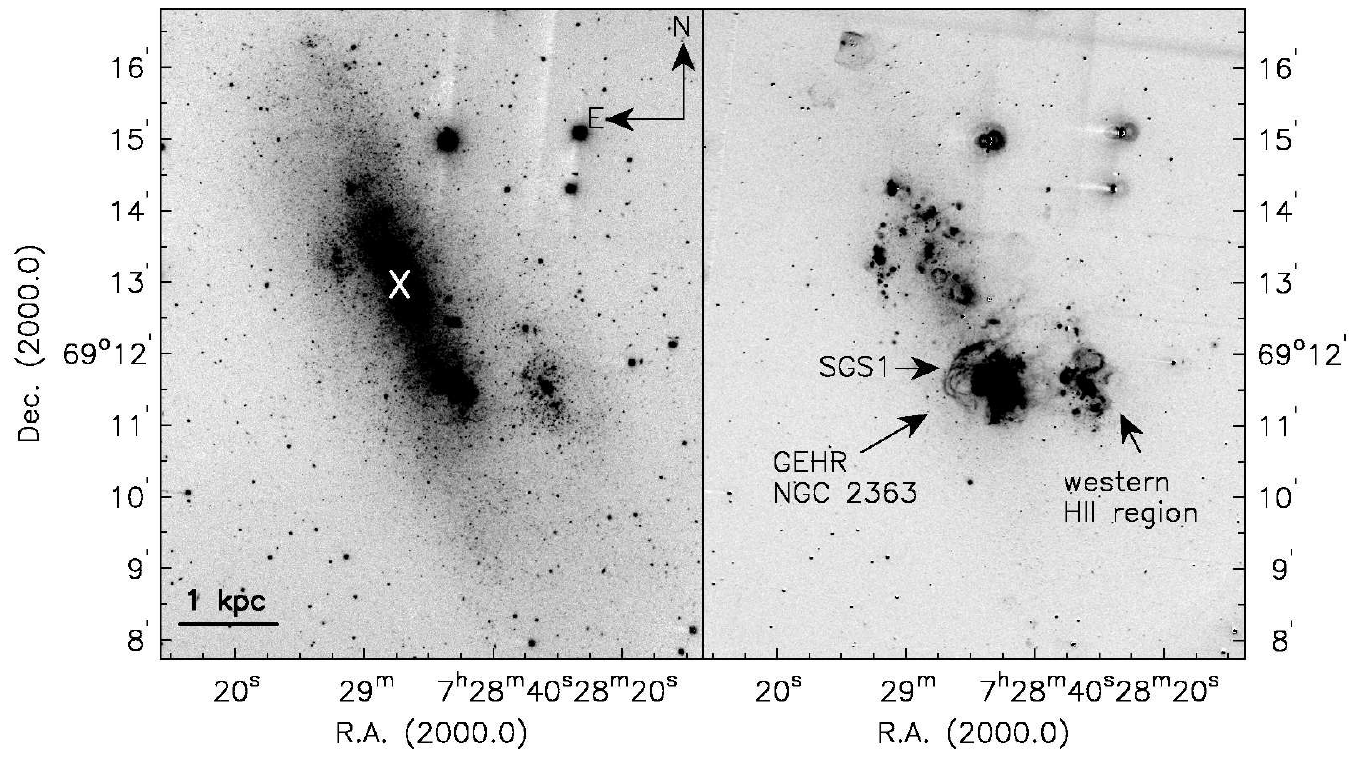}
\caption[\emph{R}-band image and continuum-subtracted \Ha\ image of
  NGC\,2366.]{\emph{R}-band image (left panel) and continuum-subtracted \Ha\
  image (right panel) of NGC\,2366. The optical centre is indicated by a white
  cross on the \emph{R}-band image. The contrast of the \Ha\ image is chosen
  in a way to emphasise the small-scale structure. The supergiant shell SGS1
  \citep[see][]{vanEymeren2007} as well as two of the main \HII\ regions are
  marked. In order to stress weaker structures and to differentiate them from
  the noise, we used adaptive filters based on the H-transform algorithm
  \citep{Richter1991}.}
\label{N2366r+ha}
\end{figure*}

Irregular dwarf galaxies are known to be the sites of giant star formation
regions. The feedback between massive stars and the interstellar medium (ISM),
and on larger scales also the intergalactic medium (IGM), is one of the most
important processes in the evolution of these galaxies. Massive stars are
strong sources of radiation and mechanical energy. Photoionisation is the most
probable mechanism, but also shocks that are produced by stellar winds and
supernova (SN) explosions inject huge amounts of energy into the ISM. This
leads to numerous ionised gas structures in and around the galactic plane,
visible on deep \Ha\ images
\citep[e.g.,][]{Hunter1997,Martin1998,Bomans1997,vanEymeren2007}. However,
ionised gas also exists at kpc distances away from any place of current star
formation \citep{Hunter1993}. These structures might be fragmented cool
($T\approx$\,10$^4$\,K) shell structures that were left behind by the
expanding hot gas \citep{MacLow1999}, but they can also be explained by
turbulent mixing between the hot and the neutral gas \citep{Slavin1993}.\\ 
In order to explain the observations, theoretical models were developed in
which the gas is shock-heated by collective SNe and accelerated into
the ISM forming a thin shell of swept-up ambient gas. Because of
Rayleigh-Taylor instabilities the shell can rupture and the hot gas is
expelled through tunnel-like features, called chimneys, into the halo of the
host galaxy \citep{Norman1989}. On the way into the halo and in the halo
itself, the gas cools down and depending on the strength of the gravitational
potential, the hot gas might fall back onto the galactic disc in the shape of
cool clouds, which is described in the \emph{Galactic Fountain} scenario
\citep{Shapiro1976}.\\
In some cases, the gas can be accelerated to velocities beyond the escape
velocity of the host galaxy. Especially the relatively low escape velocities of
dwarf galaxies will facilitate the removal of substantial amounts of
interstellar matter. Therefore, the question comes up whether the gas stays
gravitationally bound to the galactic disc (outflow) or whether it can escape
from the gravitational potential by becoming a freely flowing wind (galactic
wind). Numerical simulations by \citet{MacLow1999} model superbubble blowout
and blow-away in dark matter dominated dwarf galaxies of different mass and
with different energy input. Their models show that only in low mass
dwarfs ($M\rm _{gas}=10^6\,M_{\odot}$) a significant fraction of the hot gas
can escape from the gravitational potential. The metals however, produced by
massive stars and released during a SN explosion, have a high probability to
be blown away. The fraction is almost independent of the gas mass or the
energy input of the host galaxy.\\
Therefore, star-forming irregular dwarf galaxies provide a perfect environment
to study feedback processes and to hunt for galactic winds. A number of
studies already concentrated on the kinematics of ionised gas structures
\citep[e.g.,][]{Hunter1997, Martin1998, vanEymeren2007}, others on the
distribution and kinematics of the neutral gas
\citep[e.g.,][]{Thuan2004}. Only a few studies address both components together
in a detailed analysis. Additionally, most authors use single long-slit
spectra for the examination of the ionised gas component, which usually only
cover some parts of extended target objects. During the last decade, a lot of
progress has been made in the field of the so-called 3d spectroscopy. This
makes it possible to observe large parts of a galaxy or even the whole
galaxy together with sufficient spectral information in one exposure
\citep[e.g.,][]{Wilcots2001}.\\
In this paper we present a study of the neutral and ionised gas component in
the nearby irregular dwarf galaxy NGC\,2366. The \Ha\ emission as a tracer of
the ionised gas component was observed with a Fabry-Perot (FP) interferometer
that provides us with a complete spatial coverage of the stellar disc and
relevant spectral information. The FP \Ha\ data are complemented by optical
imaging and \HI\ synthesis data.\\
NGC\,2366 is classified as a barred Magellanic-type irregular (IB(s)m) dwarf
galaxy \citep{deVaucouleurs1991}. We adopt a distance of 3.44\,Mpc
from \citet{Tolstoy1995} that places NGC\,2366 in the M81 group. The nearest
neighbour is NGC\,2403 at a projected distance of 290\,kpc, which makes any
kind of recent interaction unlikely. Its appearance in \Ha\ is dominated by the
Giant Extragalactic \HII\ Region (GEHR) NGC\,2363 in the south-west at
07$\rm ^h$28$\rm ^m$29.6$\rm ^s$, +69$\rm ^d$11$\rm ^m$34$\rm ^s$ (see
Fig.~\ref{N2366r+ha}, right panel) with a
luminosity twice as bright as 30\,Doradus \citep{Chu1994}. Recent observations
by \citet{vanEymeren2007} revealed the existence of numerous ionised gas
structures up to kpc-size, especially close to the GEHR where most of the
current star formation activity takes place. Several of them have been shown
to expand. The expansion velocities, however, were in all cases much lower
than the escape velocities of the galaxy.\\
%
%

This paper is organised as follows: The observations and the data
reduction are described in \S~2. Sect.~3 is a comparative description of
the \Ha\ and \HI\ morphology. \S~4 presents a kinematic analysis of both
\Ha\ and \HI\ data. The results are subsequently discussed in \S~5, followed
by a short summary in Sect.~6.
%

\section{Observations and data reduction}
\subsection{Optical imaging}
A 900\,s \emph{R}-band image and a deep -- 3600\,s -- \Ha\ image of NGC\,2366
were obtained on the 16th of October 2007 with the Calar Alto 3.5m telescope
equipped with MOSCA, the Multi Object Spectrograph for Calar Alto. After the
standard data reduction performed with the software package IRAF, the
continuum image was subtracted to produce an image of the pure \Ha\ line
emission. In order to emphasise weaker structures and to differentiate them
from the noise, we used adaptive filters based on the H-transform algorithm
\citep{Richter1991}. Both images are shown in Fig.~\ref{N2366r+ha}.
\subsection{The Fabry-Perot \Ha\ data}
FP interferometry of NGC\,2366 was performed on the 1st of March 2006 with the
1.93m telescope at the Observatoire de Haute-Provence, France. We used the
Marseille's scanning FP and the new photon counting camera
\citep{Gach2002}. The field of view is 5\farcm8\,$\times$\,5\farcm8 on the
512\,x\,512 pixels of the detector and is slightly limited by the interference
filter to 5\farcm5\,$\times$\,5\farcm5, which gives a spatial resolution of
0\farcs68 per pixel. The \Ha\ line was observed through an interference filter
centred at the galaxy's rest wavelength of 6564.53\,{\AA} with a FWHM of
10\,{\AA}. The free spectral range of the interferometer -- 376\skms\ -- was
scanned through 24 channels with a sampling step of 15\skms. The final
spectral resolution as measured from the night sky lines is about 50\skms. The
seeing was between 3\arcsec\ and 4\arcsec.\\
60 cycles were observed with an integration time of 10\,sec per channel, hence
240\,sec per cycle. After removing bad cycles, a total integration time of 232
min remained. We used a neon lamp for the phase and the wavelength
calibration. The data reduction was done with the software package
ADHOCw\footnote{http://www.oamp.fr/adhoc/}, written by Jacques
Boulesteix.
\subsection{The \HI\ data}
\label{HIreduction}
We were provided with a fully-reduced data cube from ``The \HI\ Nearby Galaxy
Survey'' \citep[THINGS,][]{Walter2008}, a high spectral and
spatial resolution survey of \HI\ emission in 34 nearby galaxies obtained with
the VLA in its B, C, and D configuration. The synthesised beam size is $\rm
13\arcsec\times12\arcsec$ after using a ``natural'' weighting. The spectral
resolution is 2.6\skms. For more details see \citet{Walter2008}. We applied a
3-point Hanning smoothing to the cube, which improved the noise level from 0.6
to 0.4\,mJy\,beam$^{-1}$.\\
The moment maps were created from the un-smoothed cube by removing all
emission below a 2.5\,$\sigma$ threshold. The processing and the subsequent
analysis of the \HI\ data was performed with GIPSY\footnote{The Groningen
  Image Processing System}.
\begin{figure*}
\centering
\includegraphics[width=\textwidth,viewport= 57 94 508 632,clip=]{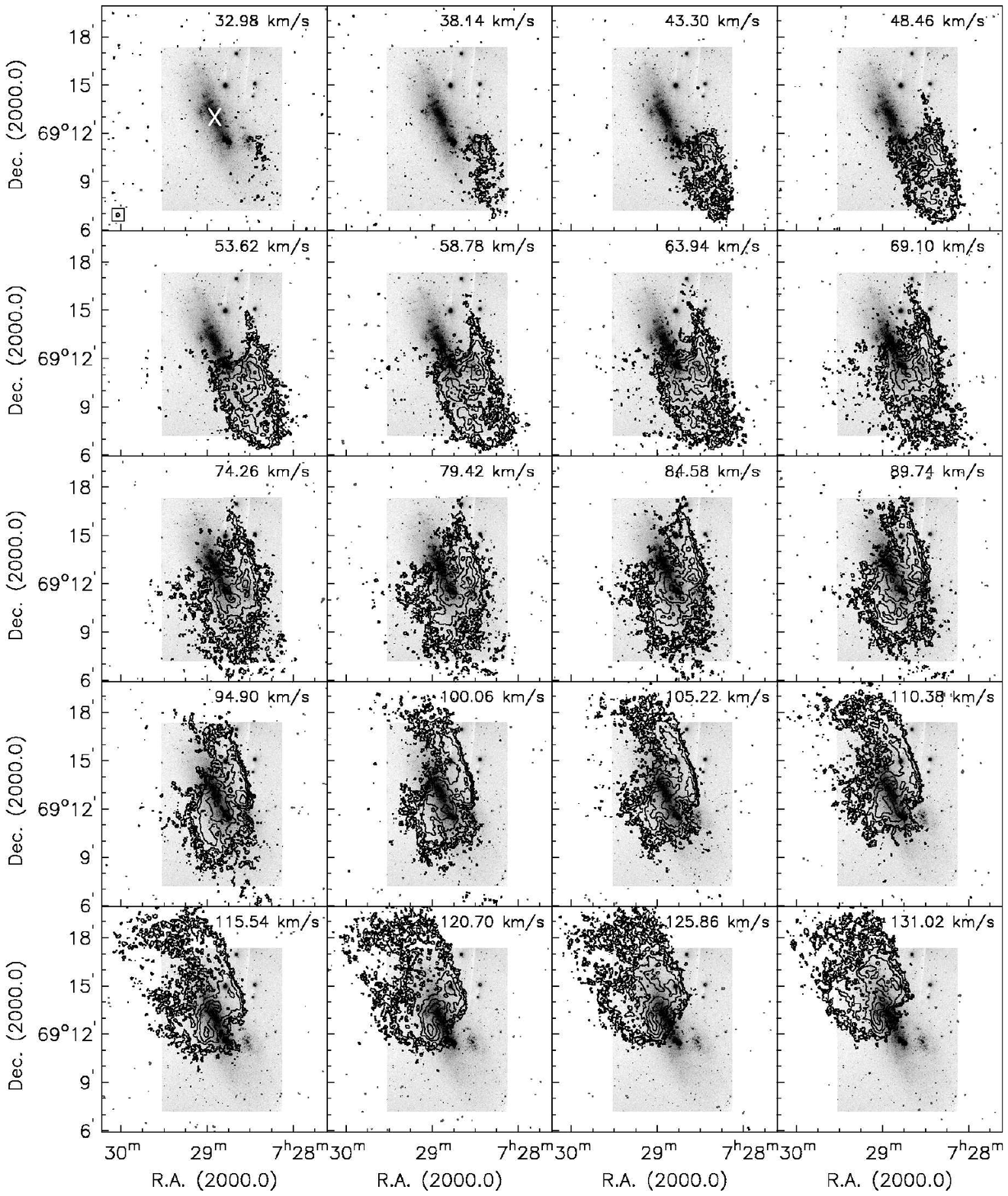}
\caption[\HI\ channel maps of NGC\,2366 (contours) superposed on the
  \emph{R}-band image.]{\HI\ channel maps of NGC\,2366 (contours) as obtained
  from the VLA using ``natural'' weighting of these data, superposed on the
  \emph{R}-band image. A 3-point Hanning smoothing was applied to the cube
  improving the noise level from 0.6\,mJy\,beam$^{-1}$ to
  0.4\,mJy\,beam$^{-1}$. The original channel spacing is 2.6\skms. Contours
  are drawn at $-$1.2 ($-$3$\sigma$), 1.2 (3$\sigma$), 3, 6, 12, and
  24\,mJy\,beam$^{-1}$. The synthesised beam is displayed in the lower left
  corner of the first channel map. The optical centre of the galaxy is marked
  by a white cross in the same channel map. The corresponding heliocentric
  velocities are printed in the upper right corner of each channel.}
\label{N2366chan}
\end{figure*}
\begin{figure*}
\centering
\includegraphics[width=\textwidth,viewport= 57 394 508 632,clip=]{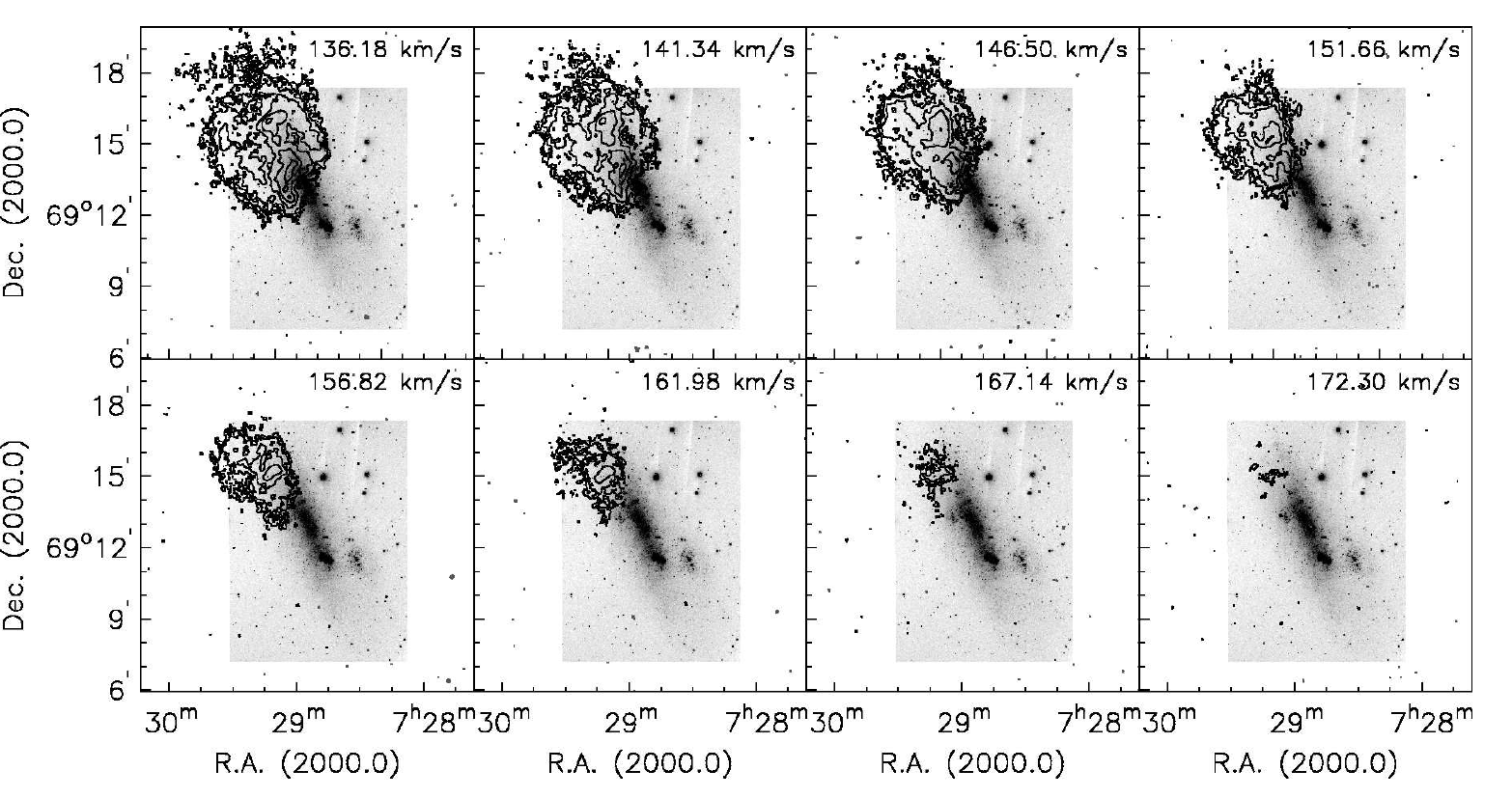}
\caption{Figure~\ref{N2366chan} to be continued.}
\label{N2366chanb}
\end{figure*}
\section{General morphology}
\subsection{Results from optical imaging}
In addition to the \emph{R}-band and the \Ha\ images presented in
Fig.~\ref{N2366r+ha}, a second grey-scale presentation of the \Ha\ image with
a different contrast in order to emphasise the weaker structures is given
in Fig.~\ref{ha}. The galaxy has a bar-like
appearance. The main star-forming complex is located at the southern end,
coinciding with the GEHR NGC\,2363 (Fig.~\ref{N2366r+ha}, right panel). The
\HII\ region west of the GEHR is sometimes referred to as a satellite galaxy
\citep[e.g.,][]{Drissen2000}. Whether an independent system or not, this \HII\ 
region is connected to the stellar disc by diffuse, filamentary gas
structures.\\
A catalogue of ionised gas structures can be found in
\citet{vanEymeren2007}. As the field of view of the \Ha\ image used by
\citet{vanEymeren2007} only covered the main part of the stellar disc, we made
some new detections on the larger image used here. To the very north of the
disc a very interesting structure was found, which is the superbubble SB4
\citep[continuing the labelling of][]{vanEymeren2007}. At the same position,
the \emph{R}-band image shows a small cluster of stars. This bubble is
remarkable because its morphology consists of a sharply edged shell pointing
to the north and a smooth shell pointing south. Additionally, there is a
second shell, SB5, heading south-east at a distance of 1\,kpc from SB4 without
an ionising star cluster nearby. Another peculiarity in NGC\,2366 are the very
small and faint \HII\ regions that are scattered along the disc on both
eastern and western side with no apparent connection to it (see Fig.~\ref{ha}).
\subsection{Results from \HI\ synthesis data}
\label{SectGMhi}
\begin{figure*}[hbt]
\centering
\includegraphics[width=\textwidth,viewport= 48 298 580 698,clip=]{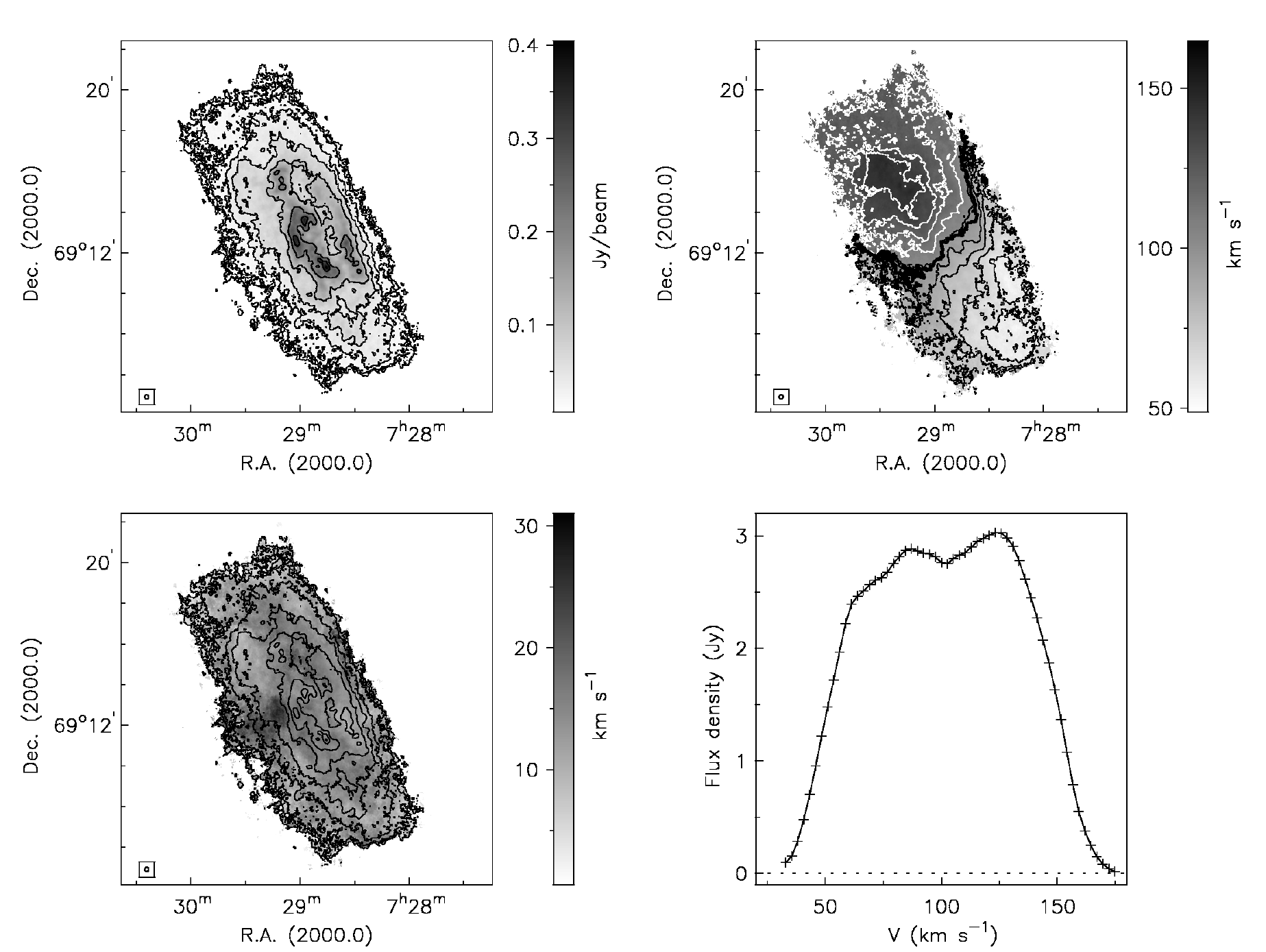}
\caption[The \HI\ moment maps of NGC\,2366.]{The \HI\ moment maps of NGC\,2366
  using ``natural'' weighting, which leads to a synthesised beam of 13\arcsec\
  $\times$12\arcsec. {\bf Top left:} The \HI\ intensity distribution (0th
  moment). Contours are drawn at 0.005, 0.01, 0.025, 0.05, 0.1, 0.2, 0.3, 0.4,
  0.5, 0.6, and 0.7\,Jy\,beam$^{-1}$ where 5\,mJy\,beam$^{-1}$ correspond to a
  column density of $\rm 3.6\times 10^{19} atoms\,cm^{-2}$. {\bf Top right:}
  The \HI\ velocity field (1st moment). Contours are drawn from 40 to
  160\skms\ in steps of 10\skms. The systemic velocity of 97\skms\ is marked
  in bold. {\bf Bottom left:} The velocity dispersion (2nd moment), overlaid
  are the same \HI\ intensity contours as on the 0th moment map. {\bf Bottom
    right:} The global intensity profile of the galaxy. The short-dashed line
  marks zero intensity.}
\label{N2366HI}
\end{figure*}

Figures~\ref{N2366chan} and \ref{N2366chanb} show the \HI\ channel maps of the
cube after applying a 3-point Hanning smoothing, superimposed on a greyscale
presentation of our \emph{R}-band image. The white cross in the first channel
marks the optical centre of the galaxy. The beam is placed into the lower left
corner of the same channel. The corresponding heliocentric velocities are
printed in the upper right corner of each channel. The \HI\ emission is very
extended (up to a factor of two in comparison to the \emph{R}-band image) and
diffuse. The most prominent structure is a tail-like feature at a velocity of
100\skms.\\
The \HI\ moment maps as well as the global intensity profile are displayed in
Fig.~\ref{N2366HI}. The \HI\ intensity distribution (upper left panel,
0th moment map) looks very patchy with several intensity maxima in the inner
parts and a more diffuse filamentary structure in the outer parts. On a larger
scale, the \HI\ forms two bright elongated structures or ridges that are
embedded in fainter and smoother gas \citep[see also][]{Hunter2001,Thuan2004}.
Both ridges are parallel to the major axis and the western one coincides with
the tail-like feature detected in the channel maps. The total \HI\ flux
density is 213\,Jy\skms, which is in good agreement with single dish
measurements by, e.g., \citet{Thuan1981}. This shows that the D array of the
VLA is very sensitive to extended emission. The \HI\ mass derived from the 0th
moment map is given in Table~\ref{TabN2366HIprop}.
\subsection{A comparison of the neutral and ionised gas distribution}
\label{SectGMcomp}
\begin{figure}
\centering
\includegraphics[width=.47\textwidth,viewport= 55 291 275 700,clip=]{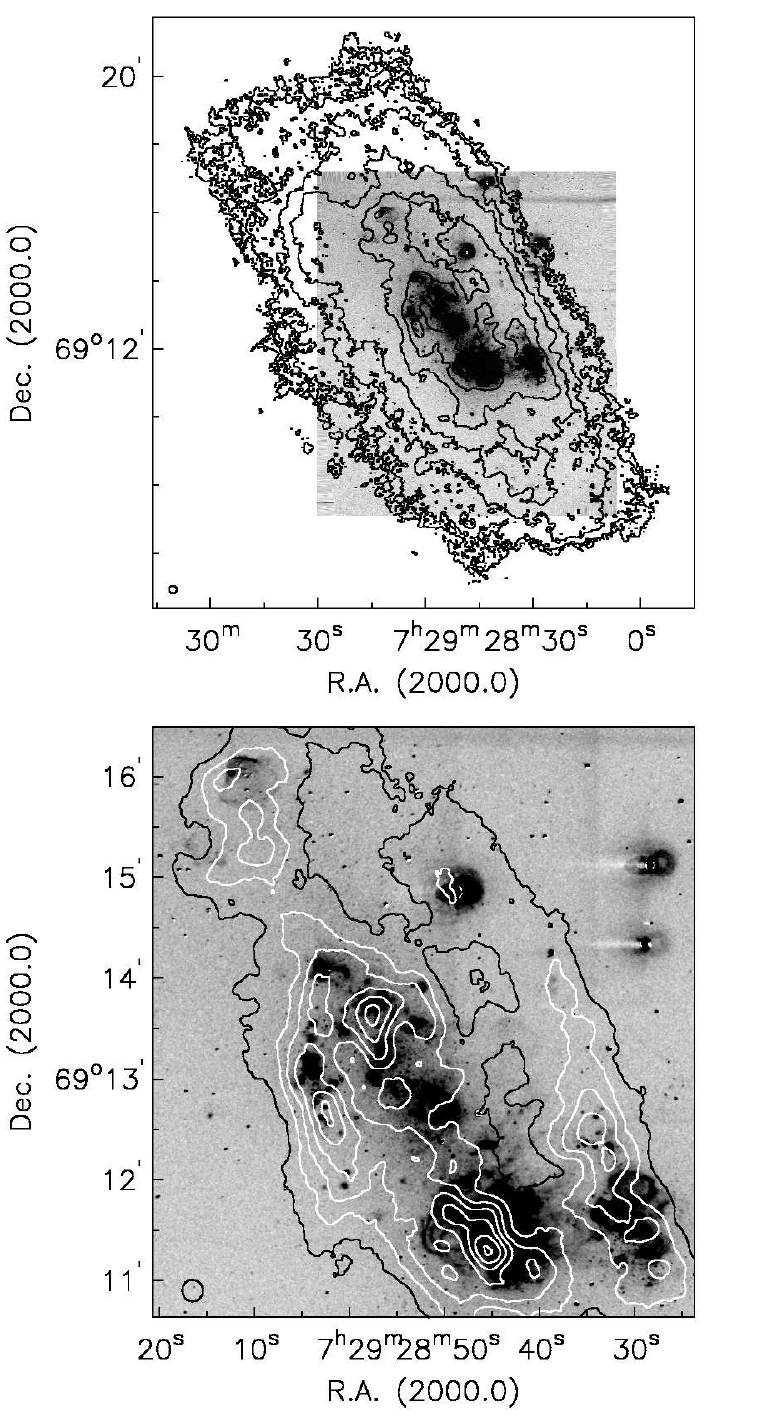}
\caption[A comparison of the \Ha\ and \HI\ morphology.]{A comparison of the
  \Ha\ and \HI\ morphology. The upper panel shows the continuum-subtracted \Ha\
  image. Overlaid in black are the \HI\ intensity contours at 0.005, 0.01,
  0.025, 0.05, 0.1, 0.2, 0.3, and 0.4\,Jy\,beam$^{-1}$. The lower panel
  displays an enlargement of the optical part. The \HI\ contour at
  0.1\,Jy\,beam$^{-1}$ is overlaid in black, the \HI\ intensity  contours from
  0.15 to 0.4\,Jy\,beam$^{-1}$ in steps of 0.05\,Jy\,beam$^{-1}$ are overlaid
  in white, which allows us to distinguish between maxima and minima in the
  neutral gas distribution.}
\label{N2366hazoom}
\end{figure}

Figure~\ref{N2366hazoom} shows a comparison of the ionised and neutral gas
distribution in NGC\,2366. In the upper panel, the \HI\ contours are overlaid
over the continuum-subtracted \Ha\ image. The optical galaxy lies in the
centre of the \HI\ distribution, i.e., in the area of highest \HI\
intensity. The \HI\ maximum flux density coincides with the GEHR, the region
of currently the strongest star formation in NGC\,2366.\\
For a more detailed study, Fig.~\ref{N2366hazoom}, lower panel shows an
enlargement of the main optical part of NGC\,2366. In order to better
differentiate between \HI\ maxima and minima, the \HI\ contour at
0.1\,Jy\,beam$^{-1}$ is overlaid in black, the \HI\ intensity  contours from
0.15 to 0.4\,Jy\,beam$^{-1}$ in steps of 0.05\,Jy\,beam$^{-1}$ are overlaid in
white. First, it can be seen that the \HI\ maximum coinciding with the GEHR is
slightly offset from the centre of the \Ha\ emission. The shift is of the
order of one beam size or even more and therefore not a resolution
effect. Such offsets have also been found in other galaxies like, e.g., IC\,10
\citep[][]{Hodge1994} and are explained by sequential star formation.\\
Additionally, the arm-like feature in the north-east as well as the western
\HII\ region coincide with an \HI\ intensity maximum. And finally, several
smaller \HI\ intensity peaks can be found in the northernmost part of
NGC\,2366, close to but offset from SB4. SB5 seems to be located in an \HI\
hole, which gives rise to the assumption that it expands into the neutral
medium by working like a snowplough. The neutral gas in front of the shell is
compressed and maybe shock-heated. In the back, a cavity of low densities
evolves visible as an \HI\ hole. This phenomenon cannot be observed in the
vicinity of SB4, but the filamentary structures north west of the GEHR are all
found in an area of low \HI\ column density (see Sect.~\ref{N2366outflow}).
\section{Kinematic analysis}
In this section, we will analyse the kinematics of NGC\,2366 in order to study
the characteristics of the gas and to explain the peculiarities in
its morphology. Spectra extracted from the \Ha\ and the \HI\ data
cubes show that both emission lines are sometimes split into several
components. Tasks which create velocity fields from a 3d data cube cannot take
line-splitting into account. Instead, line-splitting can then even lead to an
incorrect measurement of the line properties. Therefore, we performed a
Gaussian decomposition by interactively fitting the \Ha\ and \HI\ emission 
(IRAF task \emph{splot}) in order to measure the characteristics of the lines,
especially the peak velocity, and to create the velocity fields from these
values. Only detections above a 3\,$\sigma$ limit were considered. All given
velocities are heliocentric velocities measured along the line of sight.
\subsection{The \Ha\ velocity field}
\label{SectFP}
\begin{figure}
\centering
\includegraphics[width=.48\textwidth,viewport= 60 314 310 713,clip=]{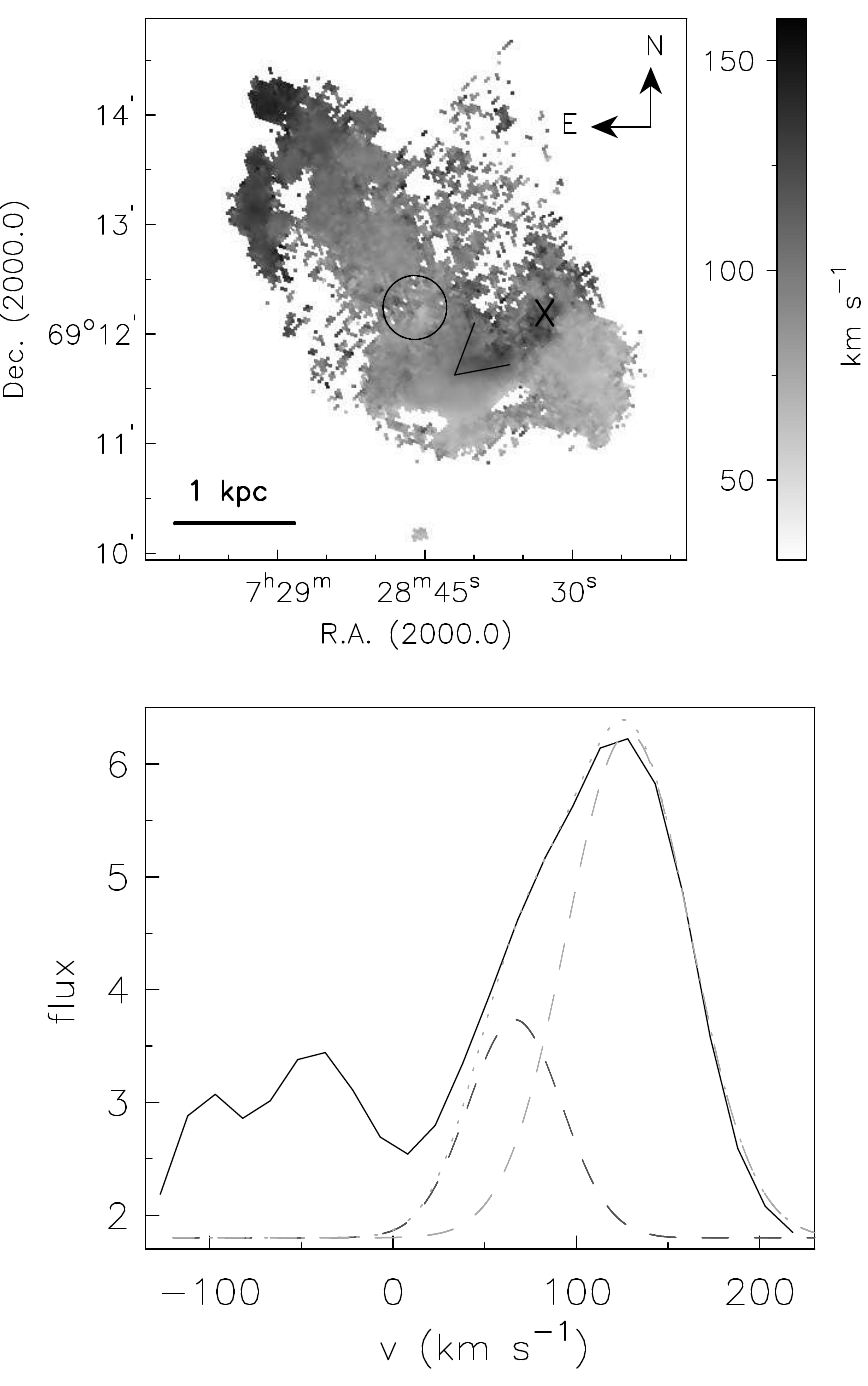}
\caption[The \Ha\ velocity field of NGC\,2366.]{The \Ha\ velocity field of
  NGC\,2366. The upper panel shows the velocity field of the
  strongest component. The positions of the two detected outflows are
  marked. The lower panel displays one example spectrum (black solid line)
  which was extracted from the FP data cube at the position of the huge
  red-shifted outflow (black cross in the \Ha\ velocity field) and which shows
  the flux in arbitrary units \emph{vs.} the velocity offset (not corrected
  for the systemic velocity of NGC\,2366). The \Ha\ line is split into a
  component at 70\skms\ (blue (dark grey) long-dashed line) and one at
  130\skms\ (red (light grey) long-dashed line). The sum of both Gaussian fits
  is plotted with a green (light grey) short-dashed line.}
\label{N2366fp}
\end{figure}

The \Ha\ velocity field (Fig.~\ref{N2366fp}, upper panel) shows the component
of the strongest intensity. In order to match the seeing and to improve
the signal to noise ratio, we summed over 3\,$\times$\,3 pixels before
performing the Gaussian decomposition. The overall velocity gradient runs from
the south-west with velocities of about 65\skms\ to the north-east with
velocities of about 120\skms. We found two major expanding gas structures
(marked in black), a large red-shifted and a faint blue-shifted outflow.\\ 
On the lower panel of Fig.~\ref{N2366fp}, we present as an example
one of the spectra extracted from the FP data cube (black solid line). The
\Ha\ line is split into two components, one at 70\skms\ (blue (dark grey)
long-dashed line) and one at 130\skms\ (red (light grey) long-dashed line). The
sum of both Gaussian fits is plotted with a green (light grey) short-dashed
line and is in good agreement with the observed
spectrum. This is one of the spectra extracted from the area of the large
red-shifted outflow (marked by a cross in Fig.~\ref{N2366fp}, upper panel). As
can be seen, the outflow dominates the \Ha\ profile, whereas the main
component is only visible as a wing on the left side. The lines to the left
side of the \Ha\ line are night sky lines which could not properly be
removed.\\
This outflow has been detected before by \citet{Roy1991} also using a FP, and
by \citet{Martin1998} and \citet{vanEymeren2007} performing long-slit echelle 
spectroscopy. \citet{Roy1991} concentrated their analysis on the GEHR
NGC\,2363, \citet{Martin1998} and \citet{vanEymeren2007} were limited by the
slit size. E.g., only one slit position by \citet{vanEymeren2007} intersects
the outflow, which made it impossible to define its whole extend. This new FP
exposure now shows the complete size of the outflow. Assuming a distance of
3.44\,Mpc, we estimated a total length of 1.4\,kpc, which is two times greater
than measured by \citet{vanEymeren2007} and even four times greater than
observed by \citet{Roy1991}. This length makes it one of the largest outflows
ever detected in a dwarf galaxy. The gas expands with a velocity of up to
50\skms\ red-shifted in comparison to the rotational gradient, which is almost
a factor of 2 higher than the detections of \citet{Roy1991} and
\citet{vanEymeren2007}. In case of the observations by \citet{Roy1991}, the
difference in velocity is most probably due to the fact that they only
detected the edge of the outflow where the velocities are lower due to its
symmetry. The slit position of \citet{vanEymeren2007} intersects the whole
outflow. Here, the discrepancy could be due to their very limited field of
view, which did not give them a true \Ha\ rotation curve, or due to their
imprecise \HI\ velocity measurements, which they used as a reference value for
outflowing gas. The comparison with the \HI\ data presented here will show
that the expansion velocity is indeed higher than measured so far (see
Sect.~\ref{N2366both}).\\
The other detected outflow is a faint blue-shifted component north of
the GEHR with velocities of 60 to 70\skms\ in comparison to the general
velocities of 90 to 100\skms, which is discussed in more detail in
Sect.~\ref{N2366outflow}.\\
As the \Ha\ line is only split in two areas across the galaxy and as the
outflow is both times the dominant component, only the main \Ha\ velocity
field is shown here.
\subsection{The \HI\ velocity field}
\label{Sec_N2366HI}
In order to analyse the general \HI\ kinematics, we worked on the moment maps
created with GIPSY (see Sect.~\ref{HIreduction}). The \HI\ velocity field (1st
moment map), displayed on the upper right panel of Fig.~\ref{N2366HI}, is
fairly regular in the inner parts, but disturbed in the outer parts. The
velocity gradient goes from the south-west with velocities of about
60\skms\ to the north-east with velocities of about 145\skms, which is similar
to the \Ha\ velocity gradient. The isovelocity contours close at both ends of
the galaxy, which might be caused by the bar and possibly by a warp structure
in the outer part. Especially the north-western part shows a completely
different kinematic behaviour to what is expected from a regularly rotating
galaxy.\\
The velocity dispersion map (Fig.~\ref{N2366HI}, lower left panel, 2nd moment)
varies between 10\skms\ in the outer parts and 17\skms\ in the inner parts of
NGC\,2366. The dispersion peaks with values of up to 28\skms\ in an area east
of the optical disc which is also clearly offset from any \HI\ maximum. Close
to the \Ha\ shell SB4 in the north, a very small \HI\ dispersion peak can be
seen with a maximum value of 25\skms, which is again offset from a nearby \HI\
maximum. This is not unusual as a high velocity dispersion suggests a high
level of turbulence which will naturally decrease the \HI\ column
densities. The high \HI\ velocity dispersions on the western and eastern edges
of the galaxy have to be treated with care. A closer look at the \HI\
velocities in these areas will show that this is an effect of low signal to
noise (see Section~\ref{N2366both}).\\
\subsection{The \HI\ rotation curve}
\begin{figure*}
\centering
\includegraphics[width=.95\textwidth,viewport= 48 299 580 693]{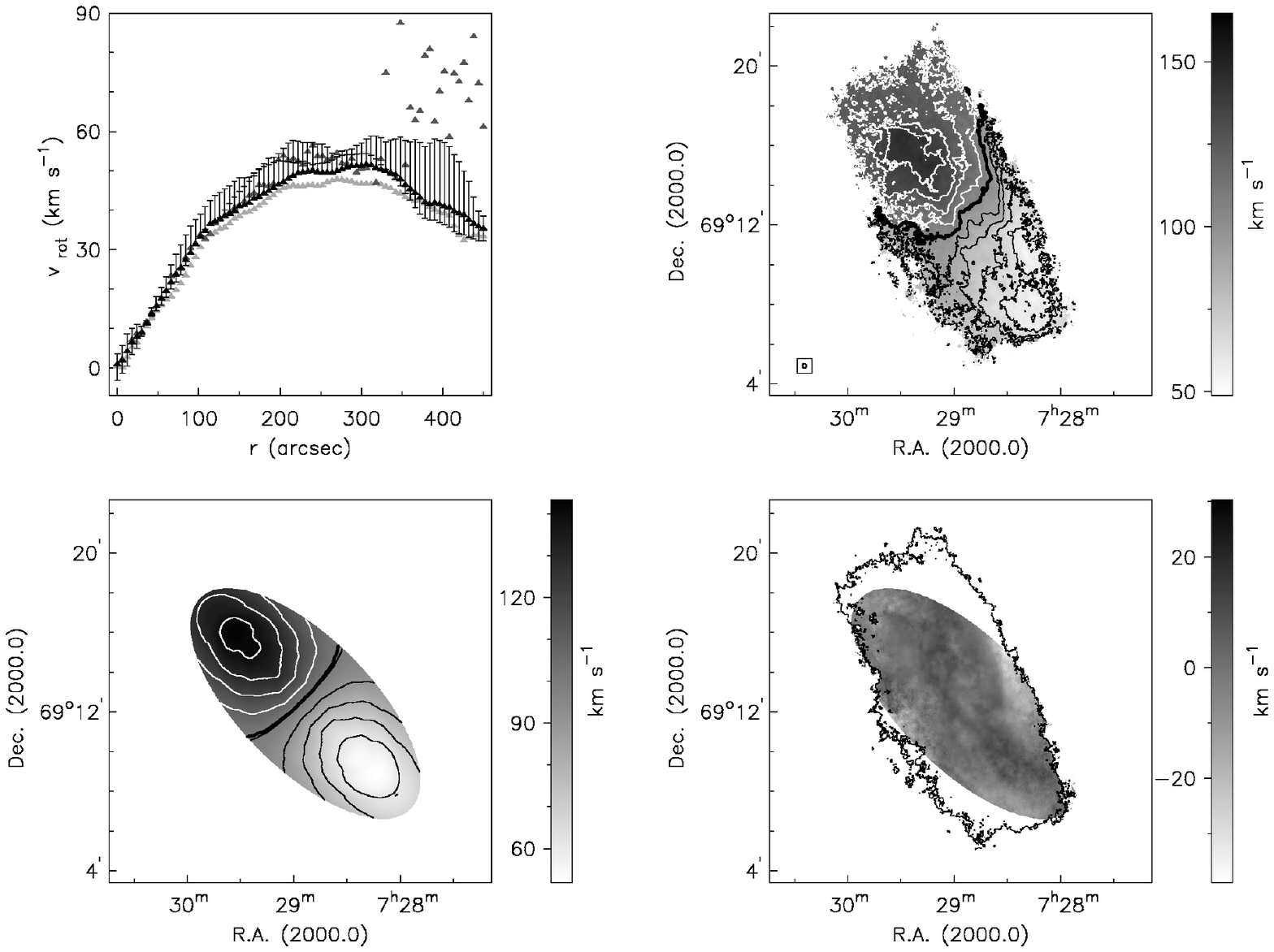}
\caption[The \HI\ kinematics of NGC\,2366.]{The \HI\ kinematics of
  NGC\,2366. {\bf Top left:} The rotation curve derived from performing a
  tilted-ring analysis. Different approaches were chosen: the black symbols
  represent the best-fitting parameters, the error bars indicate receding and
  approaching side, respectively. The green (light grey) curve was derived by
  taking the initial estimates and keep them fixed, the red (dark grey) curve
  by taking the best-fitting parameters and let them vary. {\bf Top right:}
  The observed \HI\ velocity field. {\bf Bottom left:} The model velocity
  field, based on the best-fitting parameters. {\bf Bottom right:} The
  residual map after subtracting the model from the original velocity map. The
  outermost \HI\ intensity contour at 0.005\,Jy\,beam$^{-1}$ is overlaid in
  black.}
\label{N2366rotcur}
\end{figure*}

In order to investigate the fate of the gas, we need to know some of the
kinematic parameters of NGC\,2366 like its inclination or its rotation
velocity. Therefore, we derived a rotation curve from the \HI\ data by fitting
a tilted-ring model to the observed velocity field. Figure~\ref{N2366rotcur},
upper left panel shows the resulting curves.\\
At the beginning, initial estimates for the kinematic parameters had to be
defined, which were obtained by interactively fitting ellipses to the \HI\
intensity distribution using the GIPSY task \emph{ellfit}. These were then
used as an input for the tilted-ring fitting routine \emph{rotcur}. The width
of the rings was chosen to be half the spatial resolution, i.e., 6\arcsec\ in
this case. In order to get the most precise values, three different approaches
were made by always combining receding and approaching side. First, the initial
estimates were all kept fixed. The resulting curve is indicated by the green
(light grey) symbols in Fig.~\ref{N2366rotcur}, upper left panel. In a second
approach, the parameters were iteratively defined for all rings. Up to a
radius of 300\,\arcsec, which corresponds to a distance of 5\,kpc from the
dynamic centre, no significant deviation or a sudden change of any of the
parameters was noticed so that an average value (given in
Table~\ref{TabN2366HIprop}) was taken for each parameter (black symbols). In
this case, a rotation curve was also measured for receding and approaching side
alone, indicated by the error bars of the black symbols. As a last approach,
the so derived parameters were all kept free in order to reproduce the result
of the second approach (red (dark grey) symbols).\\
The green (light grey) symbols are in very good agreement with the black ones
over the whole range of 450\,\arcsec. The red (dark grey) symbols follow the
black and green (light grey) ones up to a radius of 200\,\arcsec. Beyond a
radius of 300\,\arcsec, the deviation is of the order of 30 to 40\skms\ and
the differences between receding and approaching side become larger. The
reason for this is that the filling factors of the rings drop from about 1 to
about 0.5 at a radius of 300\arcsec\ as the galaxy does not have a clear
elliptical shape. This leads to a higher uncertainty in calculating the
rotation velocity. Therefore, all values above this radius have to be treated
with care.\\
In general, the shape of the derived rotation curve fits the ones expected for
a dwarf galaxy. In the inner 100\,\arcsec\ the velocity gradient is very steep
and linear, indicating solid body rotation, which is a characteristic sign of
dwarf galaxies. From 100\,\arcsec\ to 200\,\arcsec\ the curve is still linear,
but less steep coming to a plateau at 220\,\arcsec. From 320\,\arcsec\ on, the
rotation curve declines, which has already been implied by the closing
velocity contours in the 1st moment map (Fig.~\ref{N2366HI}, upper right
panel).\\
The best-fitting parameters derived in the iterative approach are given in
Table~\ref{TabN2366HIprop}. The systemic velocity of 97\skms\ is in good
agreement with the values measured by \citet{Hunter2001} and \citet{Thuan2004}
(99 and 101\skms, respectively). We derived an inclination of 64\degr\ and a
position angle of 45\degr\ in comparison to $i=65$\degr\ and $PA=46$\degr\
published by, e.g., \citet{Hunter2001}. Our derived rotation velocity of
50\skms\ is also in very good agreement with the values measured, e.g., by
\citet{Hunter2001} of 47\skms.\\
In order to prove the reliability of the derived parameters, a model velocity
field with the best-fitting parameters was created using the GIPSY task
\emph{velfi} (Fig.~\ref{N2366rotcur}, lower left panel) and subtracted from
the original velocity map (upper right panel). The residual map can be seen in
Fig.~\ref{N2366rotcur}, lower right panel. Overlaid
in black is the outermost \HI\ intensity contour. The overall velocity field
is well represented by our derived parameters except for the outer parts in
the north-west and south-east. Here, the residuals reach values of more than
20\skms\ in comparison to a general value of $\pm$\,10\skms. The north-western
part could not be fitted with this model.
\begin{table}
\centering
\caption{\HI\ properties measured from the THINGS data.}
\label{TabN2366HIprop}
$$
\begin{tabular}{lcc}
  \hline
  \hline
  \noalign{\smallskip}
Parameters [Unit] & NGC\,2366\\
  \hline
  \noalign{\smallskip}
  optical centre$^{\mathrm{a}}$: &\\
  ~~$\alpha$ (J2000.0) & 07$\rm ^h$ 28$\rm ^m$ 54.6$\rm ^s$\\
  ~~$\delta$ (J2000.0) & +69\degr\ 12\arcmin\ 57\arcsec\\
  $D$ [Mpc]$^{\mathrm{b}}$ & 3.44\\
\hline
\noalign{\smallskip}
  dynamic centre$^{\mathrm{c}}$: &\\
  ~~$\alpha$ (J2000.0) & 07$\rm ^h$ 28$\rm ^m$ 55.4$\rm ^s$\\
  ~~$\delta$ (J2000.0) & +69\degr\ 12\arcmin\ 27\arcsec\\
  \vsys\ [\kms]$^{\mathrm{c}}$ & 97\\
  $i$ [\degr]$^{\mathrm{c}}$ & 64\\
  $PA$ [\degr]$^{\mathrm{c}}$ & 45\\
\hline
\noalign{\smallskip}
\vrot\ [\kms]$^{\mathrm{c}}$ & 50\\
 $F_{\rm HI}$ [Jy \kms]     & 213\\
  $M_{\rm HI}$ [$\rm 10^8$\,\Msun] & 5.95\\
  \HI\ diameter [\arcmin] & 15\,$\times$\,7\\
  ~~~~~~~~"~~~~~~~  [kpc]  & 15\,$\times$\,7\\
  \HI\ / opt. ratio & 1.9\,$\times$\,2.8\\
  $<\sigma>$ [\kms] & 10 / 17\\
  $\sigma_{\rm Peak}$ [\kms] & 28.2\\
  $r_{\rm HI,max}$ [kpc] & 7.5\\
  $M_{\rm dyn}$ [$\rm 10^9$\,\Msun] & 4.3\\
  \noalign{\smallskip}
  \hline
\end{tabular}
$$
\begin{list}{}{}
\item[$^{\mathrm{a}}$] Data from NED.
\item[$^{\mathrm{b}}$] Distance measured from cepheids \citep{Tolstoy1995}.
\item[$^{\mathrm{c}}$] Derived by fitting a tilted-ring model to the \HI\ data.
\end{list}
\end{table}
\subsection{A comparison of the neutral and ionised gas kinematics}
\label{N2366both}
\begin{figure}
\centering
\includegraphics[width=.48\textwidth,viewport= 58 494 310 703, clip=]{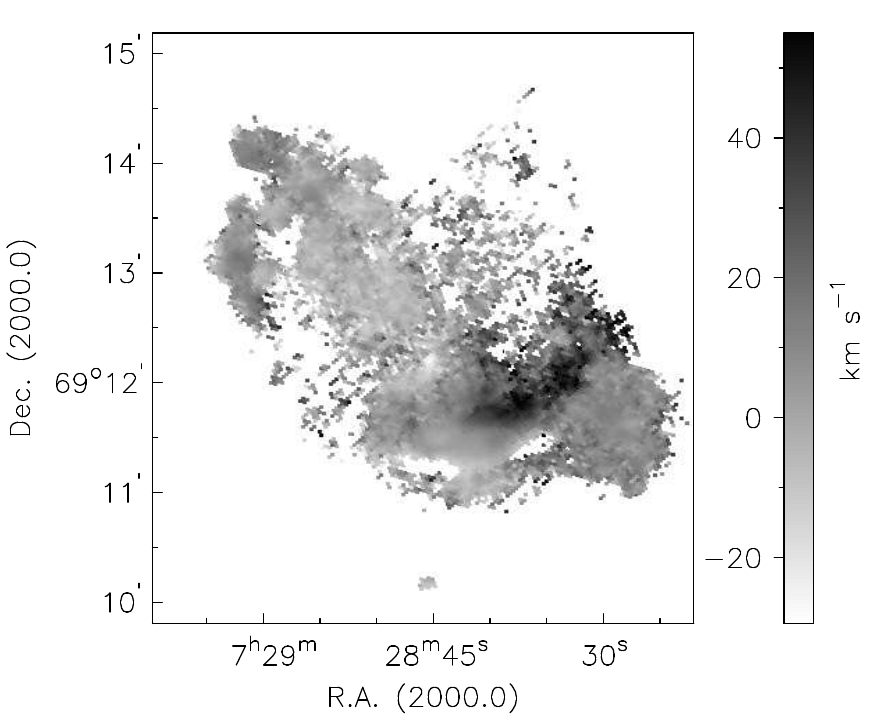}
\caption[A comparison of the neutral and ionised gas kinematics.]{A comparison
 of the neutral and ionised gas kinematics. The residuals after subtracting
 the \HI\ velocity field from the \Ha\ velocity field are shown.}
\label{N2366all}
\end{figure}
\begin{figure*}
\centering
\includegraphics[width=\textwidth,viewport= 58 524 490 708]{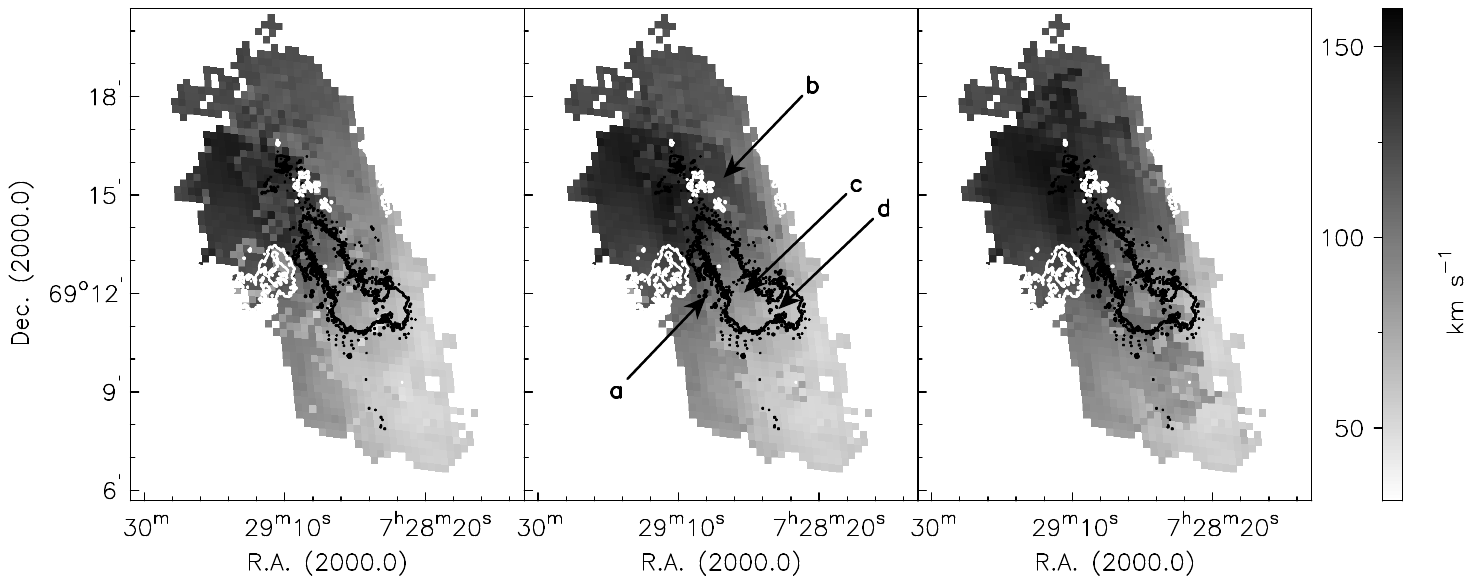}
\caption[Gaussian decomposition of the \HI\ line.]{Gaussian decomposition of
  the \HI\ line. Blue-shifted (left panel), main (middle panel) and
  red-shifted (right panel) components of the \HI\ velocities are
  shown. Overlaid in white are the \HI\ velocity dispersion contours at 20 and
  25\skms\ together with the outermost \Ha\ intensity contour in black. The
  positions of the four spectra shown in the next figure are indicated.}
\label{N2366.hi.decomp}
\end{figure*}
\begin{figure}
 \centering
\includegraphics[width=.462\textwidth,viewport=37 49 300 712,clip=]{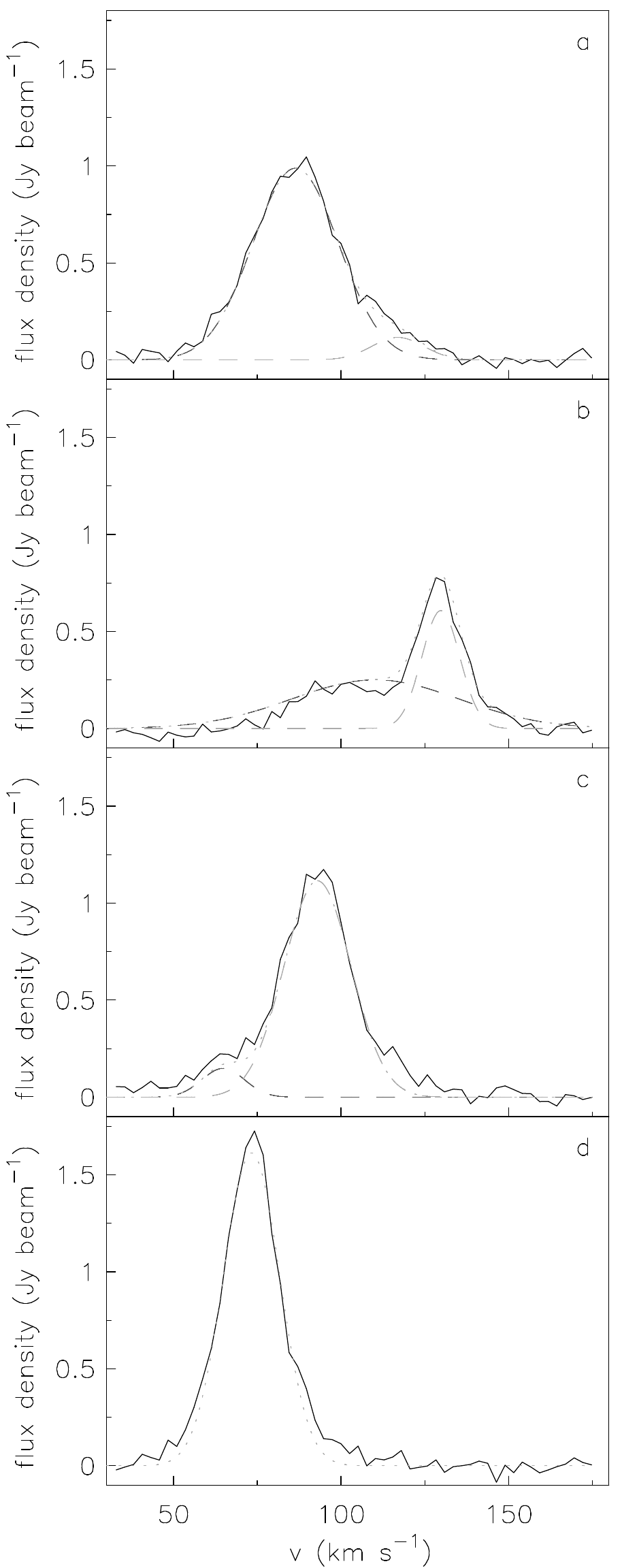}
\caption[Some examples of \HI\ line profiles.]{Some examples of \HI\ line
  profiles. The Gaussian profiles fitted to the single components are plotted
  with blue (dark grey) long-dashed lines and red (light grey) long-dashed
  lines, the resulting sum of the profiles is displayed with a green (light
  grey) short-dashed line. {\bf (a)} The eastern ridge with an additional
  red-shifted component. {\bf (b)} The western ridge with an additional
  blue-shifted component. {\bf (c)} A hint of the blue-shifted component north
  of the GEHR with $v_{\rm helio}=65$\skms. {\bf (d)} The red-shifted outflow
  north-west of the GEHR with $v_{\rm helio}=130$\skms\ is not detected in
  \HI.}
\label{n2366lineprofileshi}
\end{figure}
\begin{figure*}
\centering
\includegraphics[width=\textwidth,viewport= 45 364 500 683]{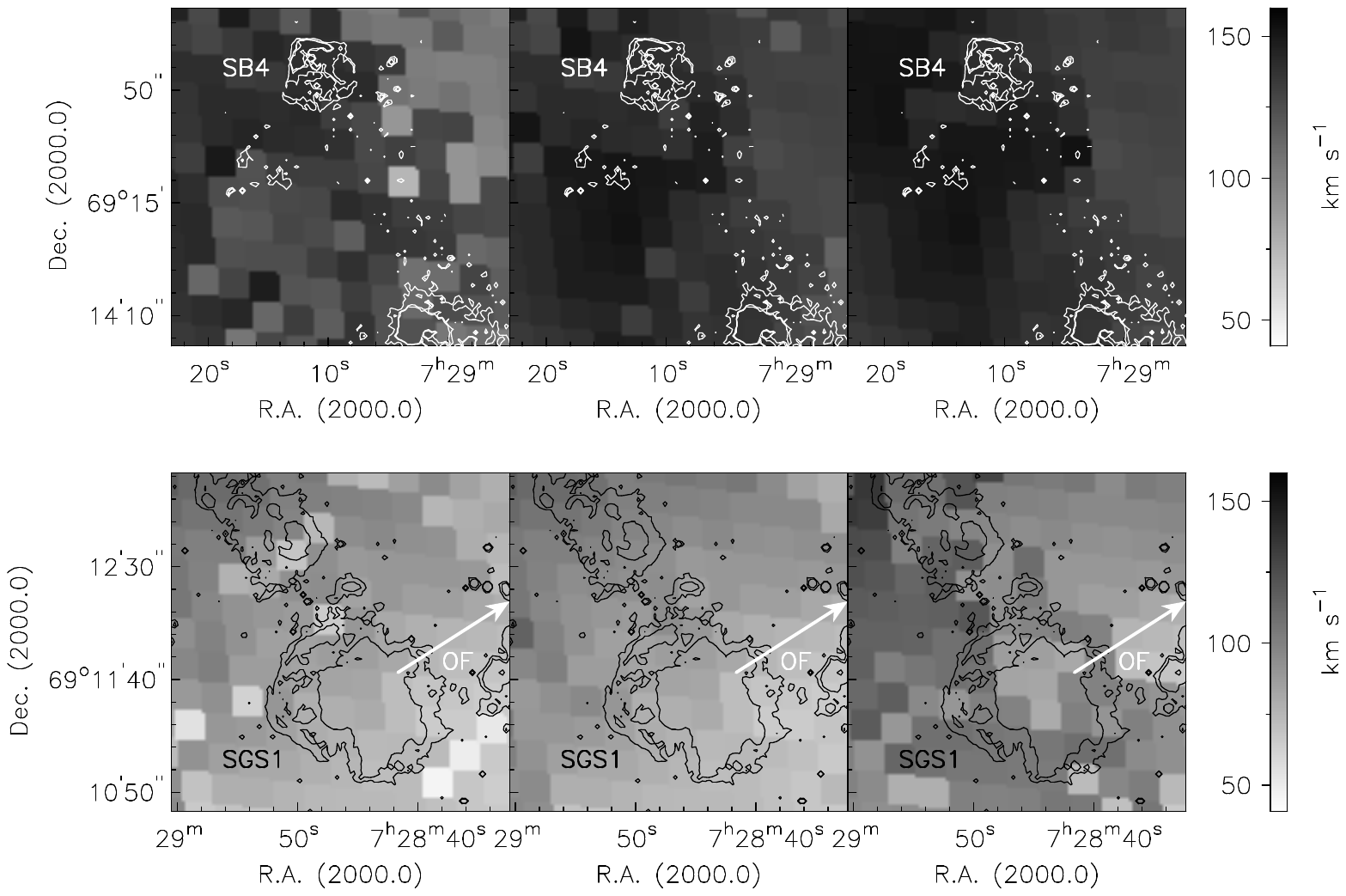}
\caption[An enlargement of the central region of NGC\,2366.]{The same as
  Fig.~\ref{N2366.hi.decomp} for two peculiar areas in the central region of
  NGC\,2366. The upper row shows the superbubble SB4 and its surroundings, the
  lower row presents the GEHR and the onset of the northern tail. Again, the
  three \HI\ velocity components are plotted. Overlaid in black are the \Ha\
  intensity contours. The superbubble SB4, the supergiant shell SGS1
  \citep[see][]{vanEymeren2007} and the outflow OF are marked.}
\label{N2366.hi+ha.decomp}
\end{figure*}

The \Ha\ velocity field (see Fig.~\ref{N2366fp}) shows two major deviations
from the overall rotation velocity, a red-shifted component north west
of the GEHR and a blue-shifted one north of the GEHR. As a next step, the \HI\
velocity field was subtracted from the \Ha\ velocity
field. Therefore, the FP data were smoothed to fit the \HI\ spatial resolution
of $\rm 13\arcsec \times 12\arcsec$. The residual map is shown in
Fig.~\ref{N2366all}. At most positions, the velocities of the neutral and
ionised gas are in good agreement with offsets of less than $\pm$10\skms. The
large red-shifted outflow clearly stands out with an expansion velocity of up
to 50\skms. Additionally, the whole northern part except for the arm-like
feature to the east seems to have a faint blue-shifted component with an
offset maximum of 30\skms\ and a median offset of 15\skms. This expanding gas
could already be seen on the original \Ha\ velocity map. It coincides with an
area of diffuse and filamentary emission between the GEHR and the northern
tail (see Fig.~\ref{N2366hazoom}, lower panel). For a detailed catalogue of
the single filaments see \citet{vanEymeren2007}, their Fig.~A.1 and Table~A1.\\
In order to look closer at the \HI\ kinematics in the area of the expanding
ionised gas and to check the regions of high \HI\ velocity dispersion, we
performed a Gaussian decomposition of the \HI\ data as described above. The
result is shown in Fig.~\ref{N2366.hi.decomp} and some example spectra
extracted from the \HI\ cube are given in Fig.~\ref{n2366lineprofileshi}
together with the fitted Gaussian profiles for the single components
(long-dashed blue (dark grey) and red (light grey) lines) and the resulting
sum (short-dashed green (light grey) lines). The velocities were
averaged over one beam size, which is in this case roughly 12\arcsec\ in both
spatial directions. The strongest component is presented in the middle
panel. Overlaid in white are the \HI\ velocity dispersion contours at 20 and
25\skms\ and the outer \Ha\ intensity contour in black. For a comparison, the
blue- and red-shifted components are shown on the left and right panel. Note
that regions where we did not find a blue- or red-shifted component were filled
with the main component.\\
Fig.~\ref{N2366.hi+ha.decomp} shows an enlargement of SB4 and the area of the
GEHR. Overlaid over the three \HI\ velocity
components are some of the \Ha\ intensity contours. The superbubble SB4, the
supergiant shell SGS1 (see Fig.~\ref{N2366r+ha}), and the outflow OF are
marked. As already mentioned in Sect.~\ref{SectGMcomp}, the huge outflow
north west of the GEHR expands into an area of low column density (a factor of
at least two lower than the surroundings). The \HI\ velocity maps in
Fig.~\ref{N2366.hi+ha.decomp}, lower row show no line-splitting, which can
also be seen by looking at a spectrum in this area
(Fig.~\ref{n2366lineprofileshi}, panel~d). Only close to the origin of the
outflow, a second component appears in the \HI\ line profile. With a velocity
of 100\skms\ in comparison to the outflow velocity of 130\skms, it is most
probably not connected to the outflow.\\
SB4 has also no counterpart in \HI\ (see Fig.~\ref{N2366.hi+ha.decomp}, upper
row). The same is true for SGS1 (lower row). Line-splitting of the \HI\
emission takes place in the north-eastern part of the GEHR going to the south,
which is at the position of the ridge mentioned by \citet{Hunter2001} or
\citet{Thuan2004}. This is discussed in more detail in Sect.~\ref{spiralarm}.\\
Finally, a hint of the faint blue-shifted component detected on the \Ha\
velocity map (see Fig.~\ref{N2366fp}) is also visible in \HI\ with the same
velocity of 65\skms\ (see Fig.~\ref{n2366lineprofileshi}, panel~c).
\section{Discussion}
Our analysis has shown that NGC\,2366 harbours many interesting features both
in its gas distribution and its gas kinematics. In the following subsections,
we try to explain the detections (and also a non-detection) and deal with the
important question what the fate of the outflows is.
\subsection{The outflows}
\label{N2366outflow}
\begin{figure}
\centering
\includegraphics[width=.49\textwidth,viewport=58 508 320 713,clip=]{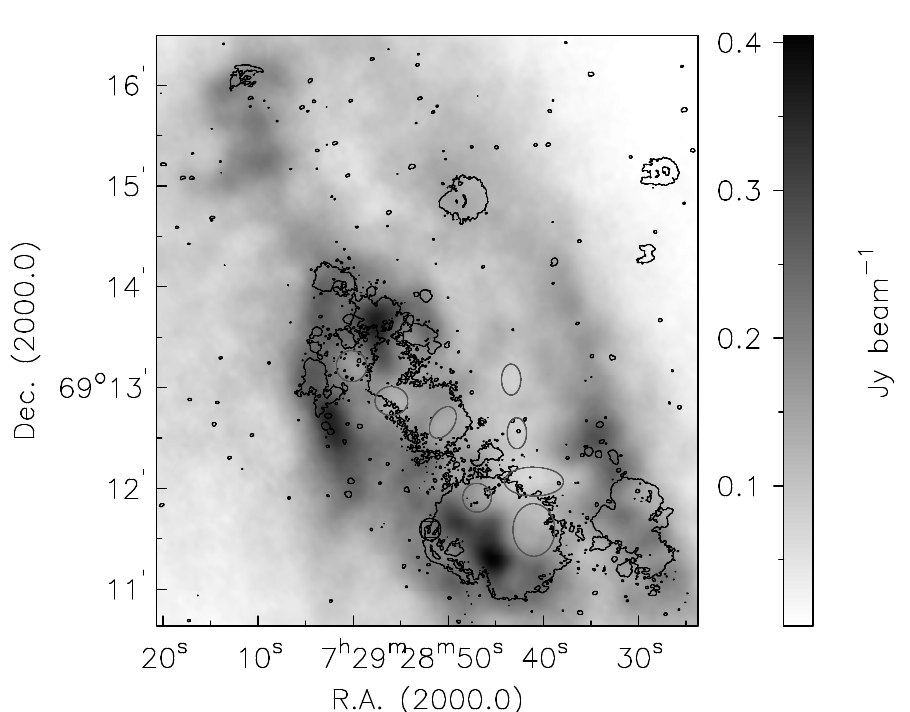}
\caption{The locations of the central \HI\ holes in NGC\,2366. The inner part
  of the \HI\ intensity distribution is shown. In order to facilitate the
  comparison of the locations of the \HI\ holes (indicated by red (light grey)
  ellipses) with the ones of the outflows, the outer \Ha\ contour is displayed
  in black.}
\label{holes}
\end{figure}

Two major outflows were found in the \Ha\ velocity field. The large red-shifted
outflow expanding from the GEHR to the north-west is not detected on the \HI\
velocity map, whereas the blue-shifted outflow north of the GEHR has a
counterpart in \HI. Additionally, we see shell-like structures (SGS1, SB4) on
the \Ha\ image with no kinematic evidence on the \HI\ map. Resolution effects
cannot be responsible for the non-detection as the structures cover a field of
several beam sizes. Unfortunately, we cannot trace the shells in the \Ha\
velocity field as SB4 lies outside the field of view and SGS1 is covered by
artificial emission caused in the detector (see
Sect.~\ref{Sect:superbubble}). However, we have already shown that SGS1 is
kinematically visible on long-slit echelle spectra \citep{vanEymeren2007}.\\
The question which now arises is why the
blue-shifted outflow is detected both in \Ha\  and \HI, whereas the
red-shifted outflow is only detected in \Ha. This is probably an effect of age
and energy input. Considering their position in the galaxy, it can be assumed
that both outflows have a different origin. The large red-shifted outflow is
most probably driven by the two large star clusters in the GEHR
\citep{Drissen2001}. Assuming a deprojected distance from the two central star
clusters of approximately 1\,kpc and a constant expansion velocity of 50\skms,
we can estimate the expansion time of the red-shifted outflow to be $\rm 2
\times\ 10^7\,yr$. According to \citet{Drissen2001}, the older star cluster
has an age of 3 to 5\,Myr. This means that the outflow must originate from a
former star formation event and is now driven further out by the new
event. The estimated cooling times for both outflows \citep[see][]{Martin1998}
lie clearly above the expansion ages so that radiative losses can be
neglected. Therefore, the gas could already have been fully ionised by the
former event. On the other hand, it is also possible that our assumption of a
constant expansion is wrong. The expansion into a dense environment could
have caused a decline in the expansion velocity of the gas over the expansion
time. A continued decline would of course have strong consequences for the
fate of the gas.\\
The blue-shifted gas probably gets its energy from a star cluster north of the
GEHR. Assuming a distance of 500\,pc and again a constant expansion velocity,
this time of 30\skms, the expansion time is comparable to the one of the huge
red-shifted outflow. As the age of this star cluster is not known, we cannot
tell whether it is alone responsible for the blue-shifted outflow. What we can
tell is that because of the comparable expansion ages, the energy input for the
blue-shifted component must have been lower or the neutral gas densities
higher, as the gas is not fully ionised, yet. Furthermore, the fact that we
can only detect a blue-shifted component does not necessarily exclude the
existence of a red-shifted component. A detection depends on the column
density. In this case, it could be possible that the blue-shifted gas runs
into an area of higher column-density, whereas the red-shifted part expands
into an area of lower column density. The same could be true for the large
red-shifted outflow. Here, we do not see a blue-shifted component in \Ha\ and
we do not see any counterpart in \HI. Possibly, the energy input was so strong
that all the gas was immediately ionised, e.g., by shock heating. The
expansion velocities are almost twice the velocities of the blue-shifted
outflow. The low \HI\ column density in comparison to its surroundings also
suggests that high amounts of neutral gas have already been ionised.\\
Both processes, ionisation of the neutral gas and blowout should affect a
smooth \HI\ intensity distribution. Figure~\ref{holes} shows the central
region of the 1st moment map. As already mentioned in Sect.~\ref{SectGMhi}, the
intensity distribution is quite patchy. At several locations associated with
filamentary and shell-like structure in \Ha, we see small holes in the \HI\
distribution indicated by red (light grey) ellipses. These \HI\ holes are
often close to the observed outflows. Especially the large red-shifted outflow
is expanding into an area of very low \HI\ column density.\\
In order to complete the discussion of the gas kinematics and the morphology,
the hot ionised gas has to be included, too. For NGC\,2366 ROSAT and XMM
Newton data are available. \citet{Stevens1998} analysed the ROSAT data which
show that if there is X-ray emission associated with NGC\,2366, it is barely
detectable. The same is true for the XMM image where we might see faint X-ray
emission in the centre of the GEHR (which is to be expected as it is the
main area of current star formation activity) as well as in the northern
part. However, no strong source associated with NGC\,2366 could be detected.
\subsection{Superbubble blowout in NGC\,2366?}
\label{Sect:superbubble}
As already mentioned in Sect.~\ref{SectFP}, \citet{Roy1991} observed NGC\,2366
with a FP interferometer centred on [O\,{\sc iii}] $\lambda$5007. Their
velocity field shows line-splitting only within the GEHR NGC\,2363. Next to
the huge outflow they also reported the detection of an expanding superbubble
around NGC\,2363 with a diameter of 200\,pc and an expansion velocity of
45\skms. Echelle spectroscopy of the central region performed by
\citet{Martin1998} and \citet{vanEymeren2007} revealed neither a blue- nor a
red-shifted component close to the centre of NGC\,2363. The bubble is also not
visible in the \HI\ velocity field (see Fig.~\ref{N2366.hi+ha.decomp}).\\
During the analysis of our FP data, we looked very carefully at the GEHR. What
we found is that not only in this but also in other galaxies, the FP
observations of bright emission line regions, e.g., of GEHRs, seem to be
affected by blurring inside the detector. This leads to an unusually high,
artificial velocity gradient in these emission line regions. In NGC\,2366,
e.g., the blurring caused a velocity gradient of more than 200\skms\ within
1\,kpc. This is unphysical in comparison to the slowly rising velocity
gradient in the other parts of the galaxy and is not observed in \HI. Assuming
that this gradient comes from true emission, its interpretation as an
expanding superbubble is indeed intimating. Therefore, we suppose that most
probably, \citet{Roy1991} interpreted an artefact as a real expanding bubble
as they were not aware of the technical problem of blurring inside the
detector. They could also have been biased by a saturation problem of the
photon counting system, leading to an artificial splitting of the observed
emission line.
\subsection{Spiral arm structure in NGC\,2366}
\label{spiralarm}
NGC\,2366 shows several peculiarities. Photometric HST observations by
\citet{Tikhonov2007} reveal an overdensity of blue stars at a galactocentric
distance of 0.9\,kpc against the overall decrease in the young-star number
density with galactocentric distance, which has also been suggested by a
\emph{B}/\emph{V} presentation by \citet{Hunter2006}. At the same distance,
our \HI\ observations as well as the ones by \citet{Hunter2001} and
\citet{Thuan2004} show the presence of two parallel ridges running along the
major axis. Another peculiarity is the \HII\ complex west of the GEHR, which
is often referred to as a satellite galaxy interacting with NGC\,2366
\citep[e.g.,][]{Drissen2000}.\\
\citet{Tikhonov2007} suggest from their results of the stellar photometry
that the \HI\ ridges, interpreted as a deprojected \HI\ ring by
\citet{Hunter2001} and \citet{Thuan2004}, could be weak spiral arms. Our
multi-wavelength study gives further evidence for the existence of two weak
spiral arms in NGC\,2366, one located at the eastern part of the stellar disc
and one at the western side (see Fig.~\ref{ha}). In \Ha, the spiral arms can be
traced by an alignment of small isolated \HII\ regions at large distances from
the disc. The most distinct ones (numbered 7 and 8) are 2.7 and 3\,kpc away
from the disc \citep[see also][]{Hunter2001}.\\
Looking again at the enlarged view of the central region
(Fig.~\ref{N2366hazoom}), we see maxima in the \HI\ column density that follow
the spiral arms out of the disc. This could indeed be
explained as a concentration of neutral gas formed by a density wave going
through the galaxy. All along the spiral arms, the \HI\ line is split into two
components, which has already been mentioned by, e.g., \citet{Thuan2004}. A
closer look at the \HI\ velocities (Fig.~\ref{N2366.hi.decomp}) shows that
next to the main component, a red-shifted component runs along the eastern arm
to the south and a blue-shifted component along the western arm to the north
(see also the example spectra in Fig.~\ref{n2366lineprofileshi}, panels~a and
b). The eastern arm crosses the supergiant shell SGS1. As the red-shifted
component is visible on large scales, it is unlikely that we can see an
additional red-shifted component belonging to the SGS1.\\
Furthermore, we compared our data to the GALEX FUV image of NGC\,2366
\citep{GildePaz2005} where the star distribution is more pronounced than on
the \Ha\ image. It also suggests the existence of two weak spiral arms,
which coincide with the \HI\ ridges and the small \HII\ regions.\\
Taking all this together, we can confirm the conclusion of
\citet{Tikhonov2007} that NGC\,2366 seems to be a weak two-armed spiral.
\subsection{Outflow or galactic wind?}
\label{outflow}
\begin{figure}
\centering
\includegraphics[width=.5\textwidth,viewport= 48 558 330 753,clip=]{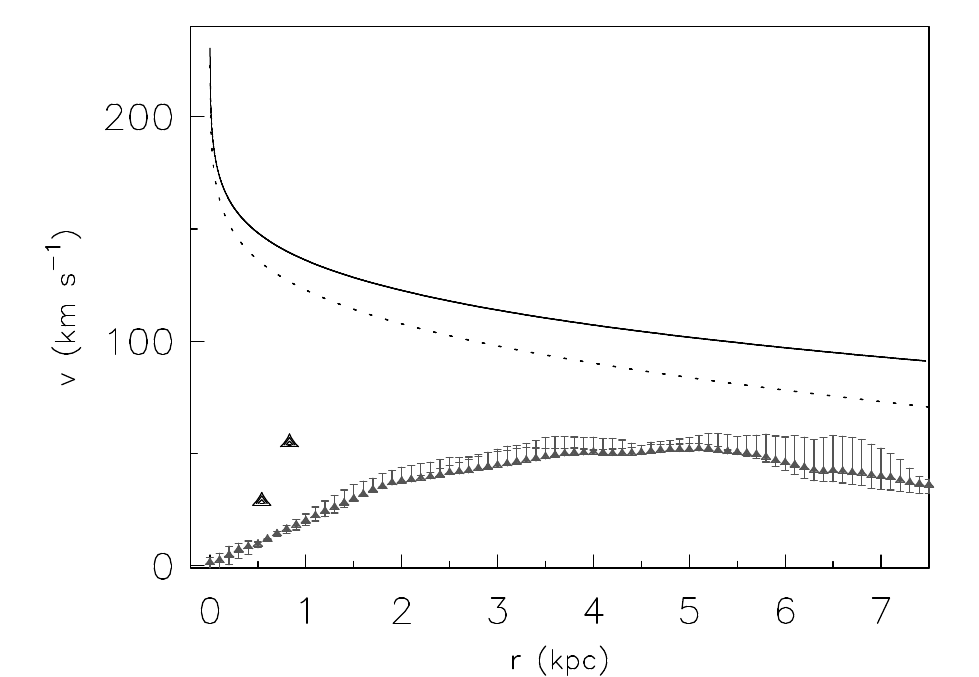}
\caption[A comparison of the escape and expansion velocities.]{Escape velocity
  for a pseudo-isothermal halo of $r_{\rm max}\rm=7.5\,kpc$ (dotted line) and
  $r_{\rm max}\rm=15\,kpc$ (solid line). The observed rotation curve is
  indicated by small grey triangles. The error bars represent receding and
  approaching side. The expanding gas structures are marked by black large
  triangles.}
\label{N2366esc}
\end{figure}

We found two main expanding gas structures in NGC\,2366. One is only
visible in \Ha\ with an expansion velocity of 50\skms, the other one shows
similar expansion velocities in \Ha\ and \HI\ of 30\skms. Both were probably
driven by a former star formation event.\\
We now want to make some statements about the fate of the gas by comparing the
expansion velocities of the outflows to the escape velocities of NGC\,2366. As
dwarf galaxies are dark matter dominated, we can calculate the escape
velocities from dark matter halo models. There are two main competing models
which describe the density distribution (and therefore the appearance of the
rotation curve) in a dark matter dominated galaxy, the NFW model derived from
Cold Dark Matter simulations \citep[e.g.,][]{Navarro1996, Moore1998,
  Moore1999} which predicts a cuspy core, and the empirically derived
pseudo-isothermal (ISO) halo \citep{Binney1987} predicting a core of constant
density. This disagreement between simulations and observations, the so-called
cusp-core discrepancy has been discussed in many publications during the last
decade \citep[e.g.,][]{deBlok2002}. Recent studies have shown that the rotation
curves of dwarf and low surface brightness galaxies can indeed better be
described by an ISO halo, at least in the inner few kpcs
\citep[][]{Kuzio2008,Spano2008}.\\
As the detected outflows are all closely located to the dynamical centre of
NGC\,2366, we decided to model the escape velocities by using the ISO
halo. This decision was further supported by a mass decomposition which
confirms that the rotation curve of NGC\,2366 can better be described by the
ISO halo \citep{vanEymeren2008PhD}. Its density profile is given by
\begin{equation}
\rho_{\rm ISO}(r)=\rho_0\left(1+\left(\frac{r}{r_{\rm c}}\right)^2\right)^{-1}
\end{equation}
with $\rho_0$ being the central density and $r_{\rm c}$ the core radius. The
escape velocity is then given by
\begin{equation}
 v_{\rm esc}(r)=\sqrt{2\,v_{\rm c}^2\left(1+\log\left(\frac{r_{\rm max}}{r}\right)\right)}
\label{equesc}
\end{equation}
with $v_{\rm c}$ being the circular velocity and $r_{\rm max}$ being the
maximum radius of the dark matter halo \citep[see][]{Binney1987}.\\
We plotted the resulting escape velocities in Fig.~\ref{N2366esc}. The \HI\
rotation curve including receding and approaching side is indicated by small
grey triangular-shaped symbols. The expanding gas structures are marked by
black large triangles. We corrected the values for an inclination of 64\degr\
as calculated in Sect.~\ref{Sec_N2366HI}, which leads to an increase in
velocity of about 10\%. The circular velocity was measured from the rotation
curve to be 50\skms\ (see Table~\ref{TabN2366HIprop}). We plotted the escape
velocity for two different radii $r_{\rm max}\rm=7.5\,kpc$ (dotted line) and
$r_{\rm max}\rm=15\,kpc$ (solid line). The lower value of $r_{\rm max}$ was
chosen to equal the size of the \HI\ outer radius. This is the outermost point
where we can directly measure a contribution of dark matter from the rotation
curve, which gives us therefore a lower limit for the radius of the dark
matter halo. Most probably, the dark matter halo is much larger so that we
chose the second value to be twice the size of the \HI\ radius (which might
still be too small). Nevertheless, a higher $r_{\rm max}$ increases the escape
velocities, which means that all radii greater than the \HI\ radius decrease
the probability of a galactic wind so that the \HI\ radius can indeed be
regarded as a lower limit for the escape velocity.\\
Fig.~\ref{N2366esc} shows that both outflows have expansion velocities that
lie clearly below the escape velocity. As the ISO halo is characterised by a
core of constant density, the escape velocities in the inner parts are even
higher than by using a rapidly decreasing NFW density profile. This result is
in good agreement with hydrodynamic simulations \citep[][see
  Sect.~\ref{Introduction}]{MacLow1999}. With a gas mass of about
10$^9$\,M$_{\odot}$ NGC\,2366 is located at the upper end of the mass
distribution simulated by \citet{MacLow1999}. This means that according to
their simulations, a blow-away of gas is very unlikely. Nevertheless, the
escape velocity drops significantly at greater distances from the dynamic
centre, which enhances the probability of a galactic wind for gas in the
halo. However, the deep \Ha\ image of NGC\,2366 shows no filamentary or
shell-like structures at several kpcs from the disc. Either the emission is
too weak to be detected or the gas has not managed so far to expand to such
large distances.
\section{Summary}
Optical images, Fabry-Perot interferometric data, and \HI\ synthesis
observations were used to get new insights into the morphology and the
kinematics of the neutral and ionised gas components in the nearby irregular
dwarf galaxy NGC\,2366. The most important results are shortly summarised
here.\\\\
In agreement with recent studies we suggest the existence of two weak spiral
arms in NGC\,2366.\\
The galaxy harbours two major outflows detected in the \Ha\ velocity
field. The large red-shifted outflow has an expansion velocity of 50\skms. The
FP data allowed us for the first time to measure the whole extent of this
outflow. With a length of 1.4\,kpc it is one of the largest outflows found so
far in a dwarf galaxy. This outflow as well as the supergiant shell SGS1 and
the superbubble SB4 do not have any counterpart in \HI\ and they all expand
into an area of low \HI\ column density. We therefore suggest that major parts
of the gas have already been ionised. The second outflow is located north of
the GEHR NGC\,2363 and expands with 30\skms\ blue-shifted. It is also visible
in \HI. Several nearby small \HI\ holes show that a fraction of the neutral
gas has already been ionised or blown away. An estimation of the expansion ages
reveals that the current star clusters are probably not the first drivers of
the outflows or that the gas is slowed down which decreases the chance for a
galactic wind.\\
We compared the measured expansion velocities with the escape velocities of
the galaxy, calculated from a pseudo-isothermal halo model. In both cases and
independent of the choice of $r_{max}$, the expansion velocities of the
outflows stay far below the escape velocity, which means that the gas is still
gravitationally bound and would need an event like a SN explosion to be
accelerated to the escape velocity. Our result is in agreement with
hydrodynamic simulations and draws therefore our attention to less massive
galaxies.
\begin{acknowledgements}
The authors would like to thank Fabian Walter and the THINGS team for
providing the reduced \HI\ data cube of NGC\,2366.\\
This work was partly supported by the Deutsche Forschungsgesellschaft (DFG)
under the SFB 591, by the Research School of the Ruhr-Universit\"at Bochum,
and by the Australia Telescope National Facility, CSIRO. It is
partly based on observations collected at the Observatoire de
Haute-Provence and at the Centro Astron$\rm\acute{o}$mico Hispano
Alem$\rm\acute{a}$n (CAHA) at Calar Alto, operated jointly by the Max-Planck
Institut f\"ur Astronomie and the Instituto de Astrof\'{i}sica de
Andaluc\'{i}a (CSIC). We made use of NASA's Astrophysics Data System
(ADS) Bibliographic Services and the NASA/IPAC Extragalactic Database (NED)
which is operated by the Jet Propulsion Laboratory, California Institute of
Technology, under contract with the National Aeronautics and Space
Administration.\\
Last but not least we would like to thank the anonymous referee for his
suggestions which helped to improve this paper.
\end{acknowledgements}
\bibliographystyle{aa}
\bibliography{9585.bw}

\hyphenation{Post-Script Sprin-ger}
\begin{thebibliography}{35}
\expandafter\ifx\csname natexlab\endcsname\relax\def\natexlab#1{#1}\fi

\bibitem[{Binney \& Tremaine(1987)}]{Binney1987}
Binney, J. \& Tremaine, S. 1987, {Galactic dynamics} (Princeton, NJ, Princeton
  University Press, 1987, 747 p.)

\bibitem[{Bomans {et~al.}(1997)Bomans, Chu, \& Hopp}]{Bomans1997}
Bomans, D.~J., Chu, Y., \& Hopp, U. 1997, AJ, 113, 1678

\bibitem[{Chu \& Kennicutt(1994)}]{Chu1994}
Chu, Y. \& Kennicutt, R.~C. 1994, ApJ, 425, 720

\bibitem[{de~Blok \& Bosma(2002)}]{deBlok2002}
de~Blok, W.~J.~G. \& Bosma, A. 2002, A\&A, 385, 816

\bibitem[{de~Vaucouleurs {et~al.}(1991)de~Vaucouleurs, de~Vaucouleurs, Corwin,
  Buta, Paturel, \& Fouque}]{deVaucouleurs1991}
de~Vaucouleurs, G., de~Vaucouleurs, A., Corwin, H.~G., {et~al.} 1991, {Third
  Reference Catalogue of Bright Galaxies} (Volume 1-3, XII, 2069 pp.~7 figs..~
  Springer-Verlag Berlin Heidelberg New York)

\bibitem[{Drissen {et~al.}(2001)Drissen, Crowther, Smith, Robert, Roy, \&
  Hillier}]{Drissen2001}
Drissen, L., Crowther, P.~A., Smith, L.~J., {et~al.} 2001, ApJ, 546, 484

\bibitem[{Drissen {et~al.}(2000)Drissen, Roy, Robert, Devost, \&
  Doyon}]{Drissen2000}
Drissen, L., Roy, J.-R., Robert, C., Devost, D., \& Doyon, R. 2000, AJ, 119,
  688

\bibitem[{Gach {et~al.}(2002)Gach, Hernandez, Boulesteix, Amram, Boissin,
  Carignan, Garrido, Marcelin, {\"O}stlin, Plana, \& Rampazzo}]{Gach2002}
Gach, J.-L., Hernandez, O., Boulesteix, J., {et~al.} 2002, PASP, 114, 1043

\bibitem[{Gil~de Paz \& Madore(2005)}]{GildePaz2005}
Gil~de Paz, A. \& Madore, B.~F. 2005, ApJS, 156, 345

\bibitem[{{Hodge} {et~al.}(1994){Hodge}, {Kennicutt}, \& {Strobel}}]{Hodge1994}
{Hodge}, P., {Kennicutt}, R.~C., \& {Strobel}, N. 1994, PASP, 106, 765

\bibitem[{Hunter \& Elmegreen(2006)}]{Hunter2006}
Hunter, D.~A. \& Elmegreen, B.~G. 2006, ApJS, 162, 49

\bibitem[{Hunter {et~al.}(2001)Hunter, Elmegreen, \& van Woerden}]{Hunter2001}
Hunter, D.~A., Elmegreen, B.~G., \& van Woerden, H. 2001, ApJ, 556, 773

\bibitem[{Hunter \& Gallagher(1997)}]{Hunter1997}
Hunter, D.~A. \& Gallagher, J.~S. 1997, ApJ, 475, 65

\bibitem[{Hunter {et~al.}(1993)Hunter, Hawley, \& Gallagher}]{Hunter1993}
Hunter, D.~A., Hawley, W.~N., \& Gallagher, J.~S. 1993, AJ, 106, 1797

\bibitem[{{Kuzio de Naray} {et~al.}(2008){Kuzio de Naray}, {McGaugh}, \& {de
  Blok}}]{Kuzio2008}
{Kuzio de Naray}, R., {McGaugh}, S.~S., \& {de Blok}, W.~J.~G. 2008, \apj, 676,
  920

\bibitem[{Mac~Low \& Ferrara(1999)}]{MacLow1999}
Mac~Low, M. \& Ferrara, A. 1999, ApJ, 513, 142

\bibitem[{Martin(1998)}]{Martin1998}
Martin, C.~L. 1998, ApJ, 506, 222

\bibitem[{Moore {et~al.}(1998)Moore, Governato, Quinn, Stadel, \&
  Lake}]{Moore1998}
Moore, B., Governato, F., Quinn, T., Stadel, J., \& Lake, G. 1998, ApJ, 499, L5

\bibitem[{Moore {et~al.}(1999)Moore, Quinn, Governato, Stadel, \&
  Lake}]{Moore1999}
Moore, B., Quinn, T., Governato, F., Stadel, J., \& Lake, G. 1999, MNRAS, 310,
  1147

\bibitem[{Navarro {et~al.}(1996)Navarro, Frenk, \& White}]{Navarro1996}
Navarro, J.~F., Frenk, C.~S., \& White, S. D.~M. 1996, ApJ, 462, 563

\bibitem[{Norman \& Ikeuchi(1989)}]{Norman1989}
Norman, C.~A. \& Ikeuchi, S. 1989, ApJ, 345, 372

\bibitem[{Richter {et~al.}(1991)Richter, Lorenz, Bohm, \& Priebe}]{Richter1991}
Richter, G.~M., Lorenz, H., Bohm, P., \& Priebe, A. 1991, Astronomische
  Nachrichten, 312, 345

\bibitem[{Roy {et~al.}(1991)Roy, Boulesteix, Joncas, \& Grundseth}]{Roy1991}
Roy, J., Boulesteix, J., Joncas, G., \& Grundseth, B. 1991, ApJ, 367, 141

\bibitem[{Shapiro \& Field(1976)}]{Shapiro1976}
Shapiro, P.~R. \& Field, G.~B. 1976, ApJ, 205, 762

\bibitem[{Slavin {et~al.}(1993)Slavin, Shull, \& Begelman}]{Slavin1993}
Slavin, J.~D., Shull, J.~M., \& Begelman, M.~C. 1993, ApJ, 407, 83

\bibitem[{{Spano} {et~al.}(2008){Spano}, {Marcelin}, {Amram}, {Carignan},
  {Epinat}, \& {Hernandez}}]{Spano2008}
{Spano}, M., {Marcelin}, M., {Amram}, P., {et~al.} 2008, MNRAS, 383, 297

\bibitem[{{Stevens} \& {Strickland}(1998)}]{Stevens1998}
{Stevens}, I.~R. \& {Strickland}, D.~K. 1998, \mnras, 301, 215

\bibitem[{Thuan {et~al.}(2004)Thuan, Hibbard, \& L{\' e}vrier}]{Thuan2004}
Thuan, T.~X., Hibbard, J.~E., \& L{\' e}vrier, F. 2004, AJ, 128, 617

\bibitem[{Thuan \& Martin(1981)}]{Thuan1981}
Thuan, T.~X. \& Martin, G.~E. 1981, ApJ, 247, 823

\bibitem[{Tikhonov \& Galazutdinova(2008)}]{Tikhonov2007}
Tikhonov, N.~A. \& Galazutdinova, O.~A. 2008, Astronomy Reports, 52, 19

\bibitem[{Tolstoy {et~al.}(1995)Tolstoy, Saha, Hoessel, \&
  McQuade}]{Tolstoy1995}
Tolstoy, E., Saha, A., Hoessel, J.~G., \& McQuade, K. 1995, AJ, 110, 1640

\bibitem[{van Eymeren(2008)}]{vanEymeren2008PhD}
van Eymeren, J. 2008, PhD thesis, Astronomisches Institut der Ruhr-Universitaet
  Bochum, Germany

\bibitem[{van Eymeren {et~al.}(2007)van Eymeren, Bomans, Weis, \&
  Dettmar}]{vanEymeren2007}
van Eymeren, J., Bomans, D.~J., Weis, K., \& Dettmar, R.-J. 2007, A\&A, 474, 67

\bibitem[{Walter {et~al.}(2008)Walter, Brinks, de~Blok, Bigiel, Kennicutt,
  {Jr.}, Thornley, \& Leroy}]{Walter2008}
Walter, F., Brinks, E., de~Blok, W.~J.~G., {et~al.} 2008, ArXiv e-prints

\bibitem[{{Wilcots} \& {Thurow}(2001)}]{Wilcots2001}
{Wilcots}, E.~M. \& {Thurow}, J.~C. 2001, ApJ, 555, 758

\end{thebibliography}
\begin{appendix}
 \section{\Ha\ image and extension of the catalogue of ionised gas structures}
\label{subappN2366}
\begin{figure*}
 \centering
\includegraphics[width=.9\textwidth,viewport= 10 20 230 240,clip=]{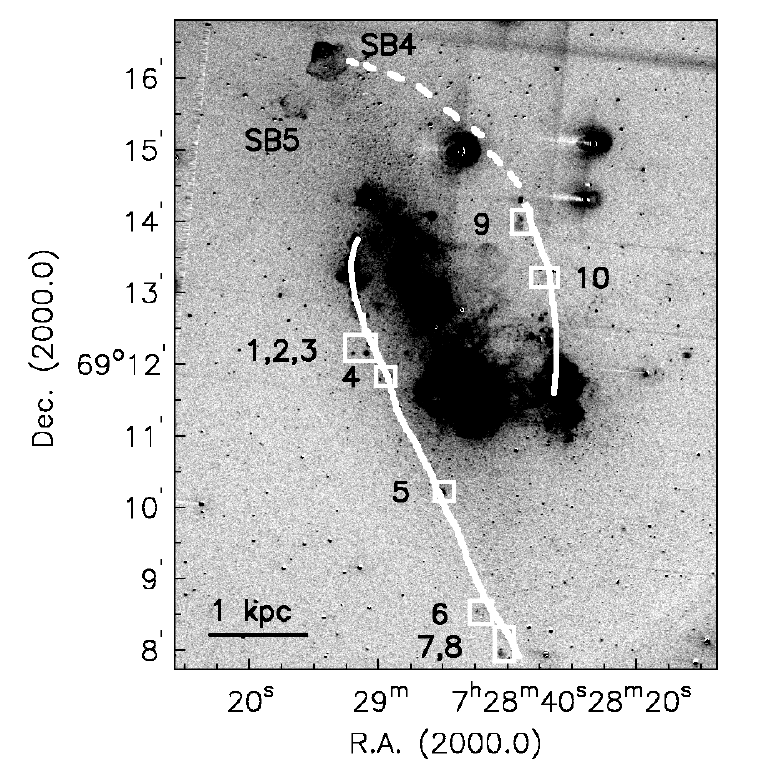}
\caption[Continuum-subtracted \Ha\ image of NGC\,2366.]{Continuum-subtracted
  \Ha\ image of NGC\,2366. The contrast is chosen in a way to demonstrate the
  large-scale structures. The extraplanar \HII\ regions are
  marked by white boxes and numbered in black. Additionally, the two northern
  superbubbles are numbered following \citet{vanEymeren2007}. The positions of
  the spiral arms are indicated by white lines.}
\label{ha}
\end{figure*}

Here, the continuum-subtracted \Ha\ image is again presented in an enlarged
version and with a different contrast as in Fig.~\ref{N2366r+ha} to emphasise
the faint and diffuse ionised gas structure (see Fig.~\ref{ha}). The
isolated \HII\ regions as well as the new detections of shell-like material
are marked and numbered. The positions of the potential spiral arms are
indicated. Additionally, an extension of the catalogue by
\citet{vanEymeren2007} can be found.
\begin{table}[h]
      \caption{The most prominent structures and their sizes in NGC\,2366 --
      an extension --.}
         \label{Filsizea}
     $$
         \begin{tabular}{lcccc}
            \hline
            \hline
            \noalign{\smallskip}
            Source & Diameter & Diameter & $v_{\rm helio}$ & FWHM\\
            & [\arcsec] & [pc] & [\kms] & [\kms]\\
            \noalign{\smallskip}
            \hline
            \noalign{\smallskip}
	    SB4 & 30 & 500 & ... & ...\\
            SB5 & 25 & 417 & ... & ...\\
            \noalign{\smallskip}
	    \hline
	    \hline
	    \end{tabular}
     $$
\end{table}
\end{appendix}
\end{document}